\documentclass[pra, 10pt, notitlepage, twocolumn]{revtex4-1}
\usepackage{graphicx}
\usepackage[caption=false]{subfig}
\usepackage{amssymb}
\usepackage{amsmath}
\usepackage{amsfonts}
\usepackage{floatrow}
\usepackage{mathtools}
\usepackage[export]{adjustbox}
\usepackage{tikz}
\usetikzlibrary{matrix,arrows,decorations.pathmorphing}

\newcommand{\ket}[1]{\left| #1 \right\rangle}

\newcommand{\proj}[1]{| #1\rangle\!\langle #1 |}

\newcommand{\Tr}{\mathrm{Tr}}
\newcommand{\Det}{\mathrm{Det}}

\newcommand{\inv}{\text{-1}}
\newcommand{\T}{\mathsf{T}}
\newcommand{\rank}{\mathrm{rank}}

\newcommand{\pref}[1]{(\ref{#1})}

\begin{document}

\title{Non-holonomic tomography II: Detecting correlations in multiqudit systems}
\author{Christopher Jackson and Steven van Enk}
\affiliation{
Oregon Center for Optical Molecular and Quantum Sciences\\
Department of Physics\\
University of Oregon, Eugene, OR 97403}

\begin{abstract}
In the context of quantum tomography, quantities called partial determinants\cite{jackson2015detecting} were recently introduced.
PDs (partial determinants) are explicit functions of the collected data which are sensitive to the presence of state-preparation-and-measurement (SPAM) correlations.
In this paper, we demonstrate further applications of the PD and its generalizations.
In particular we construct methods for detecting various types of SPAM correlation in multiqudit systems | e.g. measurement-measurement correlations.
The relationship between the PDs of each method and the correlations they are sensitive to is topological.
We give a complete classification scheme for all such methods but focus on the explicit details of only the most scalable methods,
for which the number of settings scales as $\mathcal{O}(d^4)$.
This paper is the second of a two part series where the first paper\cite{nonholo1}
is about a theoretical perspective for the PD, particularly its interpretation as a holonomy.
\end{abstract}

\maketitle

\section{Introduction}

\begin{figure}[h!]
\centering
\includegraphics[height=0.75in]{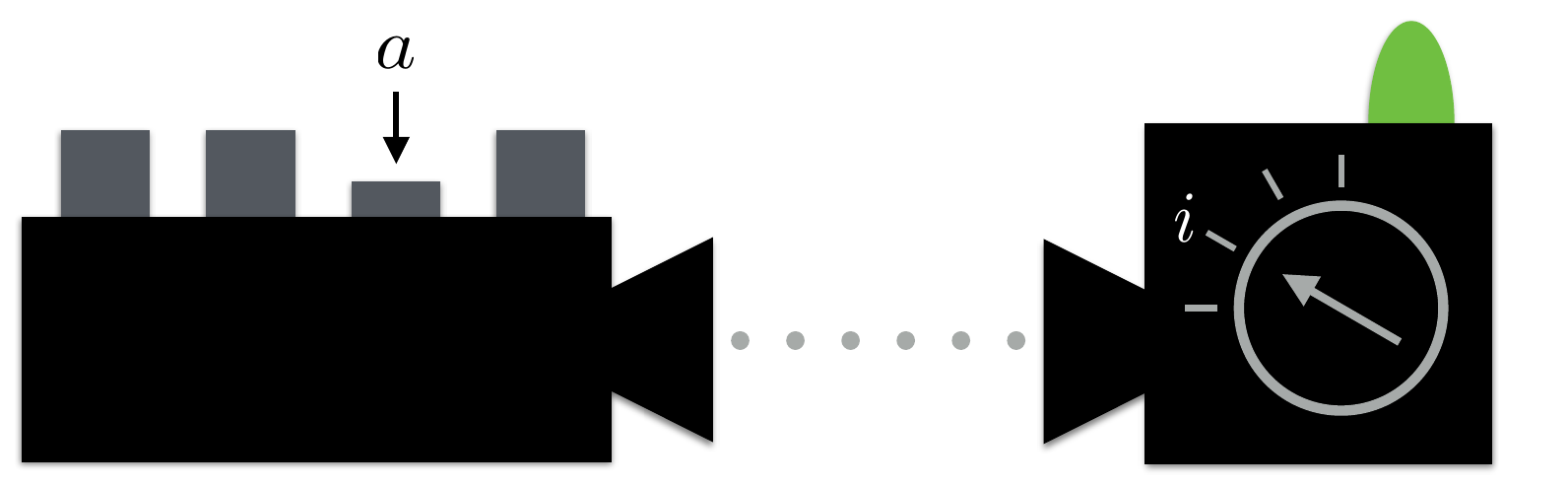}
\caption{
On the left is a device which prepares various signals on demand depending on which button, $a\in\{1,\ldots,N\}$, is pressed.
On the right is a device which blinks to indicate a signal with a certain property depending on which setting, $i\in\{1,\ldots,M\}$, a dial is turned to.
For each pair of settings $(a,i)$, if ${n_a}^i$ is the number of times the light blinks and ${N_a}^i$ is the number of times the button is pressed,
then the estimated frequency is just ${f_a}^i = {n_a}^i / {N_a}^i$.
}\label{knobs}
\end{figure}

In any quantum tomography experiment, one has the ability to perform various state preparations and measurements.
We may abstractly represent these abilities by devices with various settings (Figure \ref{knobs}.)
In standard quantum tomographies such as state, detector, or process tomography, it is assumed, respectively, that either the measurement device, the state device,
or both are already characterized and may thus provide a resource to determine the parameters associated with the yet uncharacterized devices.
Fundamental to the practice of these tomographies is a much subtler assumption:
that the performance of each device is independent of the use and history of every other device.

A problem in recent years has been the issue of estimating quantum gates
while taking into account that there are small but significant errors in the states prepared and measurements made to probe the gates, so called SPAM errors \cite{merkel}.
Any practice which takes into account SPAM errors will be generically referred to as SPAM tomography.
Several works have come out in SPAM tomography particular to the task of making estimates in spite of such conditions \cite{merkel,gst,stark},
all of which speak to the notion of a ``self-consistent tomography.''
However, these works consistently assume by fiat that the SPAM errors are uncorrelated.

In \cite{jackson2015detecting} it was demonstrated that one can test for the presence of correlated SPAM errors using so called partial determinants (PDs)
which bypass any need to estimate state or measurement parameters individually.
The logic behind the PD is simple, uncorrelated SPAM corresponds to a particular ability to factorize the estimated frequencies into a product of state and measurement parameters.
Such a factorization always exists for small enough numbers of settings but does not exist for larger numbers of settings if there are correlations.
Thus, the notion of parameter independence can be viewed as either a local or a global property.
PDs are then a measure of the contradiction that results from requiring that multiple sets of locally uncorrelated settings be consistent with each other.
In other words, SPAM correlations correspond to holonomies (or measures of global contradiction) in overcomplete tomography experiments
(hence the title, ``non-holonomic tomography.'')
Further details on this perspective may be found in \cite{nonholo1}.

For multiqudit systems, the notions of product state and product measurement introduce further kinds of factorizability in estimated frequencies.
Particularly, we will focus on systems where there is a single device associated with the preparation of multiqudit states
and a measurement device for each qudit separately (but not necessarily independently)
| i.e. systems where we expect outcome probabilities to factor into the form $\Tr \rho (E \otimes\cdots\otimes E)$.
Sure enough, PDs can be generalized to measure a degree to which such factorizations do not exist.
Thus, these generalized PDs can serve as tests for the presence of various state-state correlations, measurement-measurement correlations,
as well as mixed SPAM correlations.
Further, such generalized PDs can be much more scalable than the original PD
| $\mathcal{O}(d^4)$ settings versus $\mathcal{O}(d^{4m})$ settings where $m$ is the number of qudits.
The main portion of this work will demonstrate how to classify the various PDs one could consider.

The most basic aspect of non-holonomic tomography relies on the notion of an effectively uncorrelated system.
With this notion, one emphasizes the perspective that, although one is not able to measure individual device parameters,
correlation is simply the inability to define parameters that are organized according to a particular model.
Similar forms of analysis have come up in the context of matrix product states\cite{perez2006matrix, schon2007sequential, crosswhite2008finite},
a way of representing various kinds of many-body quantum states that is particularly elegant for calculating correlation functions.
Similar analyses can also be found in the more abstract context of (generalized) Baysian networks\cite{geiger2001stratified, garcia2005algebraic, henson2014theory} 
where the presence of hidden variables with in a model or causal structure result in a rich set of testable constraints on the probabilities associated with the observed variables.
(Bell inequalities are an example of this.)
Also from a fundamental perspective, similar analyses may be found in works on general probabilistic theories \cite{hardy2001quantum, hardy2013formalism}
where much attention is spent on the correspondence between operational descriptions of systems and the mathematical calculations that represent them.

\section{Tomography: States, Observables, and Data}

\subsection{The Born Rule Revisited}
For every quantum experiment, quantum events are counted
and the frequency of each outcome is understood to estimate the product of a state and a POVM element (Figure \ref{knobs}.)
This is the famous Born Rule, usually denoted
\begin{equation}
	{f_a}^i = \Tr\rho_a E^i
\end{equation}
where $\rho_a$ is the density operator for the state prepared according to $a$,
$E^i$ is the POVM element for an outcome of the measurement made according to $i$,
and ${f_a}^i$ is the estimated frequency.
However, we wish to consider the situation where the state-preparations and measurements (SPAM) behind these estimated frequencies actually fluctuate.
In such a case, one must modify the Born Rule to read
\begin{equation}
	{f_a}^i = {\langle\Tr\rho E\rangle_a}^i
\end{equation}
where ${\langle\rangle_a}^i$ denotes the average over the ensemble of trial runs of the devices set to $a$ and $i$ |
that is, $\rho$ and $E$ are now to be considered (positive operator-valued) random variables, distributed according to the setting $(a,i)$.

It is useful to more generally consider estimates of any statistical observable, ${S_a}^i$ such that
\begin{equation}\label{datadata}
	{S_a}^i = {\langle\Tr\rho \Sigma\rangle_a}^i,
\end{equation}
where $\Sigma$ is a Hermitian (not necessarily positive) operator-valued random variable representing the corresponding quantum observable.
The setting $i$ still represents a measurement, but can be more generally associated with a specific linear combination of outcomes which may be useful to consider
| e.g. $\Sigma^i = \proj{+_i} - \proj{-_i}$ where $\ket{\pm_i}$ are eigenstates of spin in the $i$-direction.
Any such ${S_a}^i$ will be referred to as quantum \emph{data}, calculated as the same linear combinations of measured frequencies as the observables they correspond to
| that is, ${S_a}^i = {f_a}^k{c_k}^i$ just as $\Sigma^i = E^k{c_k}^i$ for whatever ${c_k}^i$ are useful.
More traditional language would refer to ${S_a}^i$ as a ``quantum expectation value'' of the observable $i$ given state $a$.
However, for the purposes of this paper one should refrain from such language
as it is crucial to focus instead on states and observables themselves as the random variables, rather than the actual result or ``blinking of the light'' for each trial.\footnote{
The author is even inclined to suggest that states and outcomes should fundamentally be thought of as on an equal footing.}

\subsection{The Partial Determinant: A Test for Correlated SPAM errors}
In standard state, detector, and process tomographies, an experimentalist can ignore the ensemble average
because they are (respectively) able to control either the measurements, the state preparations, or both.
However, if one is doing SPAM tomography, where neither the state preparations nor measurements are assumed to be controlled,
then the ensemble average suggests the possibility that SPAM errors are correlated | i.e.
\begin{equation}\label{corrcorr}
	{\langle\Tr\rho \Sigma\rangle_a}^i \neq \Tr \langle\rho\rangle_a\langle\Sigma\rangle^i.
\end{equation}
From the perspective of doing any of the standard tomographies, this is an awkward statement indeed
because one does not have the resources necessary to access quantities such as $\langle\rho\rangle_a$ or $\langle\Sigma\rangle^i$ individually.\footnote{
One could parse correlations into two separate kinds of independence:
The first kind being when ${\langle\Tr\rho \Sigma\rangle_a}^i = \Tr {\langle\rho\rangle_a}^i {\langle\Sigma\rangle_a}^i$.
The second kind being when ${\langle\rho\rangle_a}^i = {\langle\rho\rangle_a}$ and ${\langle\Sigma\rangle_a}^i = {\langle\Sigma\rangle}^i$}
One may thus be tempted to conclude that correlations (or lack thereof) cannot be determined without access to the individual expectation values.
However, this is not the case.

Correlations such as Equation \pref{corrcorr} can be determined without individual expectation values
because equations like ${\langle\Tr\rho \Sigma\rangle_a}^i = \Tr \langle\rho\rangle_a\langle\Sigma\rangle^i$ express a very special factorizability of the data, Equation \pref{datadata}.
We thus proceed with the following operational definition:
we say that data ${S_a}^i$ is \emph{effectively (SPAM) uncorrelated} when we can express it as a simple matrix equation:
\begin{equation}\label{eff}
	{S_a}^i = {P_a}^\mu{W_\mu}^i.
\end{equation}
The rows of $P$ and columns of $W$ (when they exist) represent the states and observables, $\rho_a = {P_a}^\mu\sigma_\mu$ and $\Sigma^i = \sigma^\mu {W_\mu}^i$, 
where $\{\sigma_\mu\}$ is some operator basis and $\{\sigma^\mu\}$ is the corresponding dual basis.
Repeated indices are to be summed over.
We relax the requirement that the rows of $P$ correspond to positive operators.

Three important observations should be made about this definition.
First, Equation \pref{eff} requires that the sum on $\mu$ be over $\le n^2$ operators for an $n$-dimensional Hilbert space.
(Indeed, the notion of effectively uncorrelated is always relative to an assumed dimension, $n$.)
Second, one can always write an expression like Equation \pref{eff} (with the sum on $\mu$ being over $\le n^2$)
so long as the number of state settings, $N$, and detector settings, $M$, are both $\le n^2$.
Third, when $P$ and $W$ exist, they are in general not unique
because one could just as well use $PG$ and $G^\inv W$ where $G$ is an $n^2 \times n^2$ real invertible matrix.
The components of $G$ are gauge degrees of freedom which have been referred to as \emph{SPAM gauge}\cite{merkel,gst,stark} or \emph{blame gauge}\cite{jackson2015detecting}.

The combination of these observations suggest the following generic protocol for quantifying correlations:
First, perform SPAM tomography with $N > d^2$ and $M > d^2$.
Such SPAM tomography may be referred to as overcomplete 
because such numbers of setttings would correspond to overcomplete standard tomographies, if only the appropriate devices were controlled and well characterized.
Second, consider $d^2 \times d^2$ submatrices of the data, which can be thought of as corresponding to minimally complete tomographies.
Each such submatrix is effectively uncorrelated and thus may be associated with a ``local'' gauge degree of freedom.
Finally, check whether the states and observables for each minimally complete submatrix can be chosen so that such choices among all submatrices are consistent with each other.
It turns out that the amount of inconsistency can be quantified rather elegantly by what has been called a partial determinant.\cite{jackson2015detecting}

Such protocols can be understood as an organization of the data into a \emph{fiber bundle}.
Fiber bundles are mathematical structures which support the notions of \emph{connection} and \emph{holonomy} which are rather ubiquitous concepts.
In \cite{nonholo1}, it is demonstrated how tomography and partial determinants can be interpreted as connections and holonomies, respectively.
However, for the sake of those who are interested exclusively in potential applications, an effort has been made to avoid such language in this current paper.
Nevertheless, occasional references will be made and terms will be defined which allude to this perspective.
It is from these perspectives that the title ``Non-holonomic tomography'' is fully justified as a name for partial determinants. 

Forgetting this alternative perspective and using standard linear algebraic considerations,
instead one can make the observation that the definition for effective independence is equivalent to the statement that the data must be such that $\mathrm{rank}(S) \le n^2$.
For example, consider devices whose numbers of settings are $M = N = 2n^2$ and organize the data as
\begin{equation}\label{square}
S
=
\left[
\begin{array}{cc}
	A & B\\
	C & D
\end{array}
\right]
\end{equation}
where each corner is an $n^2 \times n^2$ matrix.
One can define a partial determinant (or PD) for this arrangement of the data,
\begin{equation}\label{PD}
	\Delta(S) = A^\inv B D^\inv C.
\end{equation}
The PD has the property that it is equal to the identity matrix if and only if the data is effectively SPAM uncorrelated.\cite{jackson2015detecting}
The proof of this is simple if one observes that $\rank(S) \le n^2$ if and only if there exist $n^2 \times n^2$ matrices $P_1$, $P_2$, $W_1$, and $W_2$ such that
\begin{equation}
S
=
\left[
\begin{array}{c}
	P_1\\
	P_2
\end{array}
\right]
\left[
\begin{array}{cc}
	W_1 & W_2
\end{array}
\right].
\end{equation}

\subsection{Multiqudit Correlations: SPAMs and Non-Localities}\label{recallterms}

In this paper we consider extensions of the general notion of a PD to multiqudit systems ($n = d^m$ for $m$ qu$d$its)
with a concentration on PD constructions with a number of device settings which scales to lowest order, $\mathcal{O}(d^4)$.
Specifically, we will focus on systems where the preparation of a multiqudit state can be represented by a single device and the measurement of each qudit can be represented separately by separate (but not necessarily independent) devices.
Uncorrelated measurements between different qudits will be referred to as \emph{local} measurements.\footnote{
This second meaning of the word ``local''  should not be too confusing as it will be clear from context
whether we are considering individual qudit observables or small numbers of state and measurement settings.}
If all measurements are effectively local and SPAM uncorrelated, then any data collected for $m$ qudits can be factored into the form
\begin{widetext}

\begin{equation}
	S_a^{ijk\ldots} = \Tr\big(\rho_a\;\Sigma_1^i\otimes\Sigma_2^j\otimes\Sigma_3^k\cdots\big) = R_a^{\lambda\mu\nu\ldots}{W_{1\lambda}}^i{W_{2\mu}}^j{W_{3\nu}}^k\cdots
\end{equation}

\end{widetext}
where $\rho_a = R_a^{\mu\nu\cdots}\sigma_\mu\otimes\sigma_\nu\otimes\cdots$ and $\Sigma_q^i = W^i_{q\mu}\sigma^\mu$ for each qudit $q \in\{1,\ldots,m\}$,
for some operator bases $\{\sigma_\mu\}$ and dual bases $\{\sigma^\mu\}$
and a sum over repeated greek indices (from $1$ to $d^2$) is always implied.\footnote{
Technically, we should write $(\sigma_q)_\mu$ to emphasize that the qudit measurements do not necessarily share a reference frame,
but we will not write this here for the sake of reducing index clutter.}

Two kinds of indices can be distinguished in these expressions.
There are those indices which correspond to device \emph{settings} ($a$, $i$, $j$, ...) which are associated with degrees of freedom which can be controlled.
Such indices may be referred to as ``external'' because they correspond to degrees of freedom outside of the quantum system being probed.
Then there are those indices which correspond to device \emph{parameters} ($\mu$, $\nu$, ...)
and represent the model | the Born rule with $d$-dimensional Hilbert spaces.
These indices may be referred to as ``internal'' because they are always summed over and thus are accompanied with gauge degrees of freedom.

For such multiqudit systems, there are now multiple kinds of correlation one can have.
We refer to correlations between states and measurements on qudit $q$ as $\text{SPAM}_q$ correlations.
Further, let us refer to correlations between measurements on qudit $q$ and measurements on qudit $p$ as $(q,p)$-nonlocalities.
Such correlations are to be understood with the notions of \emph{effectively} $\text{SPAM}_q$ uncorrelated data and \emph{effectively} $(q,p)$-local data.
Then one may proceed to categorize the various ways a partial determinant may be constructed to test if a system is $\text{SPAM}_q$ correlated or $(q,p)$-nonlocal.
For simplicity, we are only considering 2-point correlations in this paper (see Conclusions and Discussion.)

\section{Local Measurements of Two Qudits}

\begin{figure}[h!]
\centering
\includegraphics[height=0.75in]{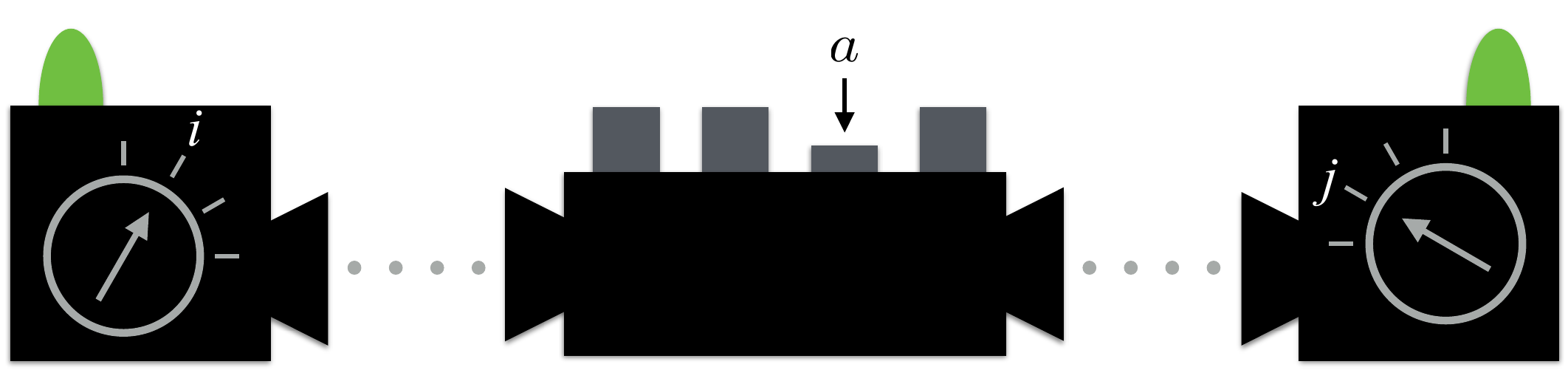}
\caption{
A two qudit experiment where there is a single device which prepares qud$^2$it states and two devices which make qudit measurements.
We would like to know if the data can be modeled by equation \pref{2qu} | that is, does the factorization in Equation \pref{2qu} exist for the accessible $S_a^{ij}$ of this experiment?
}\label{2qudev}
\end{figure}

For two qudits, a.k.a. a qud$^2$it, the data is an object with 3 (external) indices, 1 for state preparations and 2 for the measurements on each qudit.
If there are no correlations, then we may write
\begin{equation}\label{2qu}
	S_a^{ij} = R_a^{\mu\nu}{V_\mu}^i{W_\nu}^j.
\end{equation}
One can consider this as a matrix equation in the most obvious way:
\begin{equation}\label{snip1}
	{S_a}^I = {R_a}^M{X_M}^I
\end{equation}
where $M = (\mu,\nu)$, $I = (i,j)$, and ${X_M}^I = {V_\mu}^i{W_\nu}^j$, treating the two qudit measurements as one qud$^2$it measurement.
This separation of parameters suggest the original protocol \cite{jackson2015detecting} for detecting what will now be called \emph{generic} SPAM correlations,
constructing a partial determinant for $n = d^2$.

One can also consider equation (\ref{2qu}) as a matrix equation in another way:
\begin{equation}\label{snip2}
	{S_A}^j = {P_A}^\nu {W_\nu}^j
\end{equation}
where $A = (a,i)$ and ${P_A}^\nu = R_a^{\mu\nu}{V_\mu}^i$.
One can interpret this separation as the measurement settings of one qudit being used to effectively prepare states for the other qudit.
In this case, we know that if these effective-states, set by $A$, are uncorrelated with the other qudit measurements, set by $j$, (so that we may write Equation \pref{snip2})
then a smaller ($n = d$) PD of ${S_A}^j$ must be the identity.
We should stress that taking the inverse of a matrix like ${S_A}^j$ is very different from taking the inverse of a matrix like ${S_a}^I$ even if they consist of the same entries, only organized differently.

\begin{figure}[h!]
\centering
\includegraphics[height=0.75in]{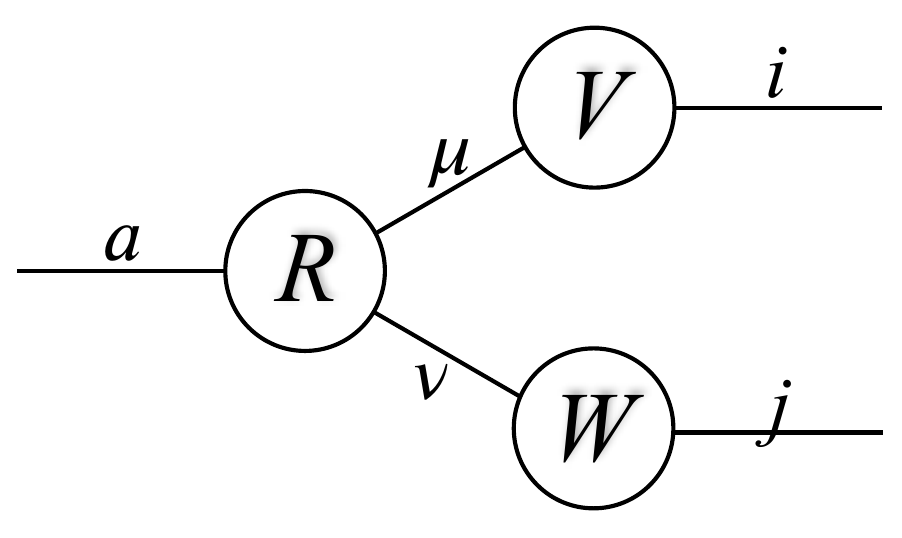}
\caption{Diagrammatic representation of effectively completely uncorrelated data, Equation \pref{2qu}.
Each internal line represents a sum over $d^2$ operators while each external line represents a setting.}\label{pic1}
\end{figure}

One sees that there are already two distinct ways to be effectively uncorrelated:
The first is to be SPAM uncorrelated in the generic sense, such that Equation \pref{snip1} exists.
In this case, the rank of ${S_a}^I$ must be $\le d^4$, particularly for $>d^4$ state settings, $a$, and $>d^4$ measurement settings, $I$.
The second is to be uncorrelated such that Equation \pref{snip2} exists.
In this case, the rank of ${S_A}^j$ must be $\le d^2$, particularly for $>d^2$ effective state settings, $A$, and $>d^2$ measurement settings, $j$, for the second qudit.

\begin{figure}[h!]
\centering
\includegraphics[height=0.7in]{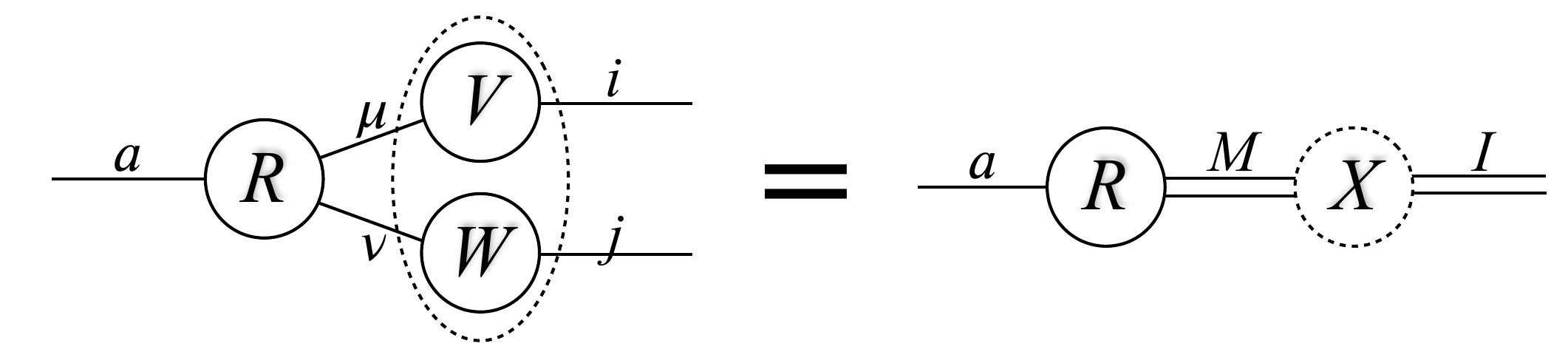}
\caption{Diagrammatic representation of Equation \pref{snip1}, a weaker form of effective independence.
The right-hand matrix has a rank bounded by $d^4$ because the dotted separation cuts two internal lines.  Double lines represent product indices.}\label{pic2}
\end{figure}

\begin{figure}[H]
\centering
\includegraphics[height=0.65in]{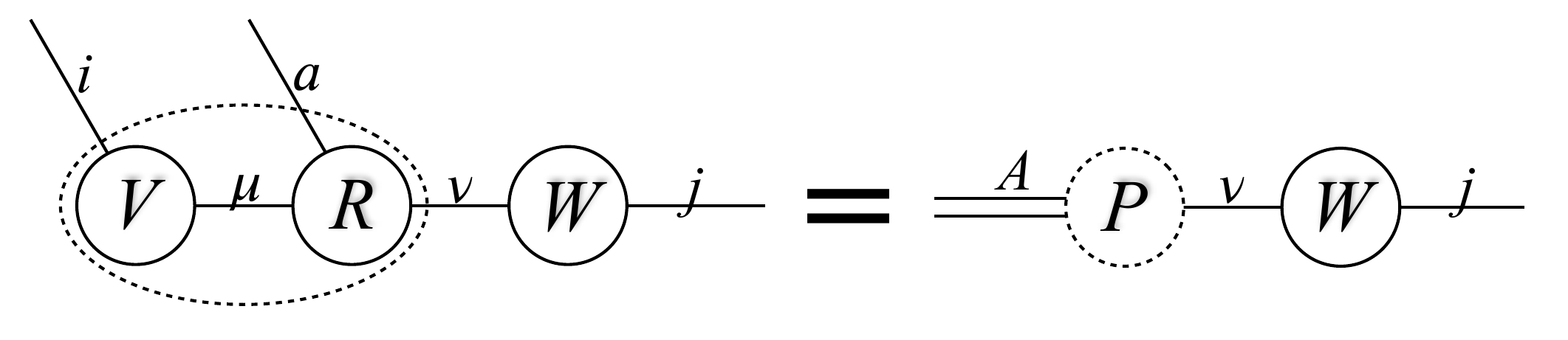}
\caption{Diagrammatic representation of Equation \pref{snip2}, another kind of weaker effective independence.
The rank of the right-hand matrix is bounded even lower by $d^2$ because the dotted separation cuts only one internal line.}\label{pic3}
\end{figure}

Equations \pref{snip1} and \pref{snip2} represent weaker forms of effective independence than Equation \pref{2qu}.
One should think of them as potential factorizations of the data which may or may not exist.
It is helpful to represent Equations \pref{2qu} through \pref{snip2} diagrammatically, as in Figures \ref{pic1} thru \ref{pic3}.
Being able to factorize the data as in Equation \pref{2qu} means that the system can be considered completely uncorrelated.
Being able to factorize the data as in Equation \pref{snip1} means that the data is effectively $\text{SPAM}_1$ and $\text{SPAM}_2$ uncorrelated.
(Recall the terminology from the end of section \ref{recallterms})
Being able to factorize the data as in Equation \pref{snip2} means that the data has effective $\text{SPAM}_2$ independence and (1,2)-locality.
Similarly, there is another factorization that results from permuting the qudits whose existence would mean the system is effectively $\text{SPAM}_1$ uncorrelated and (1,2)-local.

\subsection{Classifying Different PDs}\label{a_whole_bunch}

PDs which test for generic SPAM correlations are relatively straight forward as there are only 2 main variations on their construction.
In contrast, there are 11 distinct PDs one can consider relative to factoring the data as in Equation \pref{snip2}.
These PDs differ in their construction by the number of settings used for $a$ and $j$ (or $i$) and by how these settings are organized.
These different constructions are sensitive to different kinds of correlation.
Specifically, each PD will be equal to the identity when the system is effectively uncorrelated in a corresponding way.
In order to organize the description of these various constructions, we must establish a few definitions and some notation.

The procedure for constructing a PD can be summarized in two steps.
The basic goal is to organize the data so that it is of the same form as Equation \pref{square}.
In addition to the original PD construction, rows and columns may now be products of multiple settings.
The first step is then to organize the settings so to construct a \emph{corner} template.
The second step is to ``displace'' four instances of that corner which can then be connected in a \emph{loop}, as in Equation \pref{PD}.
This constructed matrix of four corners shall be called a \emph{square}.

\subsubsection{Generic SPAM Correlations}\label{gen}

For detecting generic SPAM correlations, such that Equation \pref{snip1} does not exist, we denote the various numbers of experimental settings by $[N:M_1,M_2]$,
where $N$ is the number of state settings (the range of $a$) and $M_q$ is the number of local measurement settings for $\text{qudit}_q$ (the range of $i$ or $j$.)
The colon can be thought of as representing the dotted separation of Figure \ref{pic2}.
Settings to the left of the colon are to be organized as a row index while settings to the right are to be columns.

To calculate a partial determinant in this case, one needs to consider corners that are $d^4 \times d^4$ which further requires $N = d^4$ and $M_1=M_2=d^2$.
(Recall that this is because the data must have rank $\le d^4$ if Equation \pref{snip1} exists.)
A square can then be assembled from such a corner in two ways, which we denote simply by multiplying the appropriate setting number by 2:
\begin{equation}
	[2d^4:2d^2,d^2] \hspace{30pt}\text{and}\hspace{30pt} [2d^4:d^2,2d^2].
\end{equation}
Thus we have two kinds of generic PD.

Importantly, we use a `2' in our bracket notation as if to suggest implementing twice as many settings, as done originally.
However, one could just as well make a square from any number of rows and columns, each $> d^4$.
In Appendix \ref{plus1}, we demonstrate how to construct an $r \times r$ PD for an $(r+1)\times(r+1)$ matrix.
Nevertheless, we will always write `$2$'s in our bracket notation for simplicity.
To summarize, factors of $d$ in this square bracket notation represent an organization template for the settings in each corner,
while `2's represent which device settings one changes when going from one corner to the next.
The nature of these representations should become much clearer in the following, more intricate factorization problem.

\subsubsection{Nonlocalities and $\text{SPAM}_q$ Correlations}

For detecting correlations such that Equation \pref{snip2} does not exist, we denote the numbers of settings by $[N;L :M]$.
We now make a distinction between $L$, the number of observable settings for $\text{qudit}_1$, used to effectively prepare states,
and $M$, the number of observable settings for $\text{qudit}_2$, used to measure them.
Again, we can interpret the colon as the dotted separation in Figure \ref{pic3}, between effective state preparations and measurements of $\text{qudit}_2$.
A semicolon after the first argument is just to distinguish the first argument as the number of (joint) state preparations.
Of course, there are actually 2 distinct schemes of type $[N;L :M]$ depending on which qudit we consider part of the effective state preparation.
We denote the other by $\pi[N;L :M]$ where $\pi$ means `permute the two qudits.'

\begin{table}[h!]
\vspace{-10pt}
\begin{equation}
\begin{array}{c|cc}
\text{Corners} & \multicolumn{2}{c}{\text{Squares}}\\
{[N;L :M]} & [2N;L :2M] & [N;2L :2M]\\\hline
{[d^2 ;1 :d^2]} & [2d^2;1 :2d^2] & [d^2;2 :2d^2]\\
{[d;d:d^2]} & [2d;d :2d^2] & [d;2d :2d^2]\\
{[1;d^2:d^2]} & [2;d^2 :2d^2] & [1;2d^2 :2d^2]
\end{array}
\end{equation}
\vspace{-10pt}
\caption{Each row is a way to make a corner while each column is a way to make a square.}\label{6table}
\end{table}

\begin{figure}[p]
\subfloat{
	\includegraphics[height=1.0in]{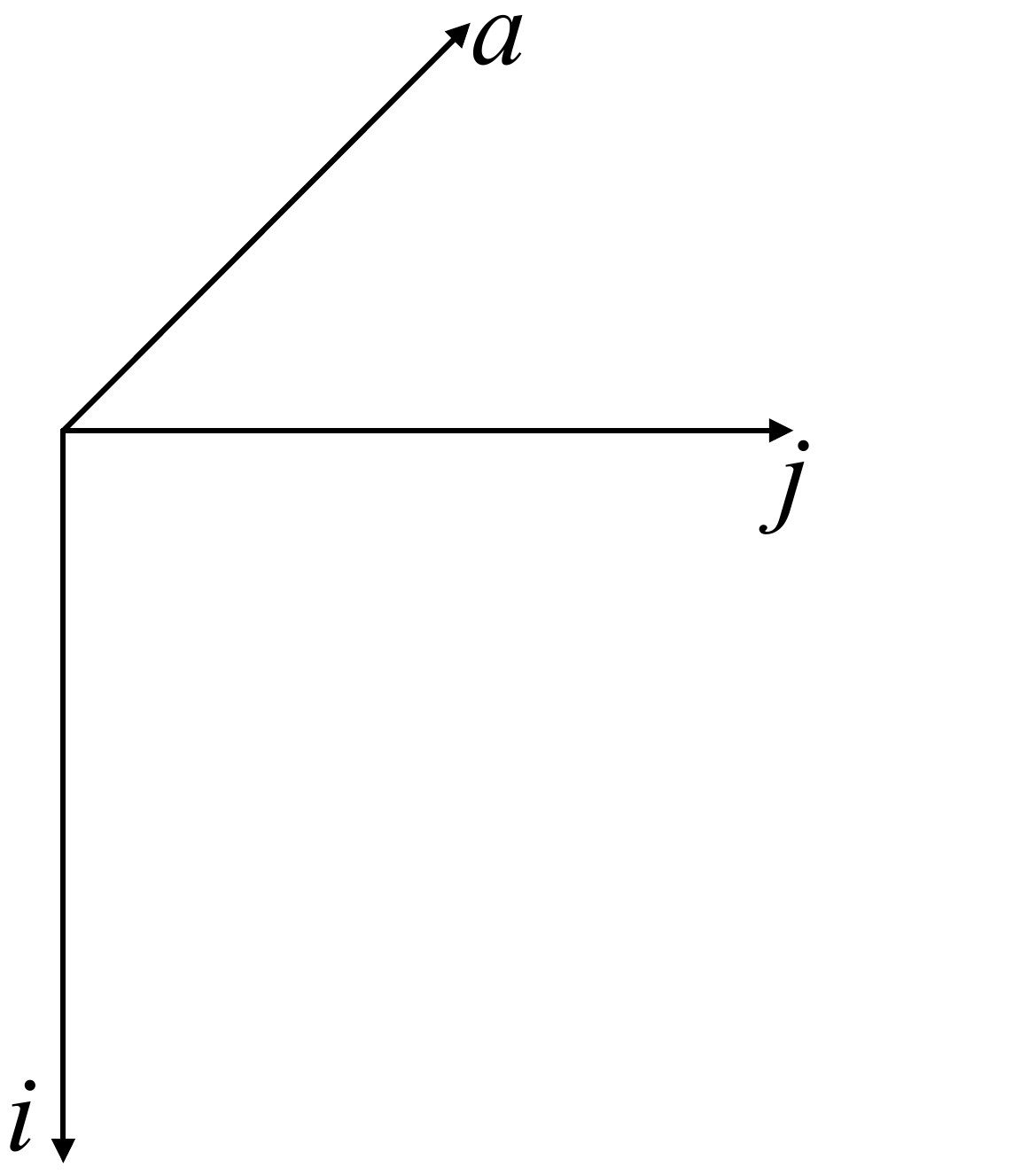}}\\
	\setcounter{subfigure}{0}
\subfloat[$\mathbf{[2d^2;1:2d^2]}$
	Using $2d^2$ states, 1 observable for $\text{qudit}_1$,
	and the usual $2d^2$ observables for $\text{qudit}_2$.
	If the $\text{qudit}_1$ observable is the identity, this is simply $\mathrm{SPAM}_2$ tomography.
        \label{end}
	]{
	\makebox[0.45\textwidth]{
	\includegraphics[height=1.15in]{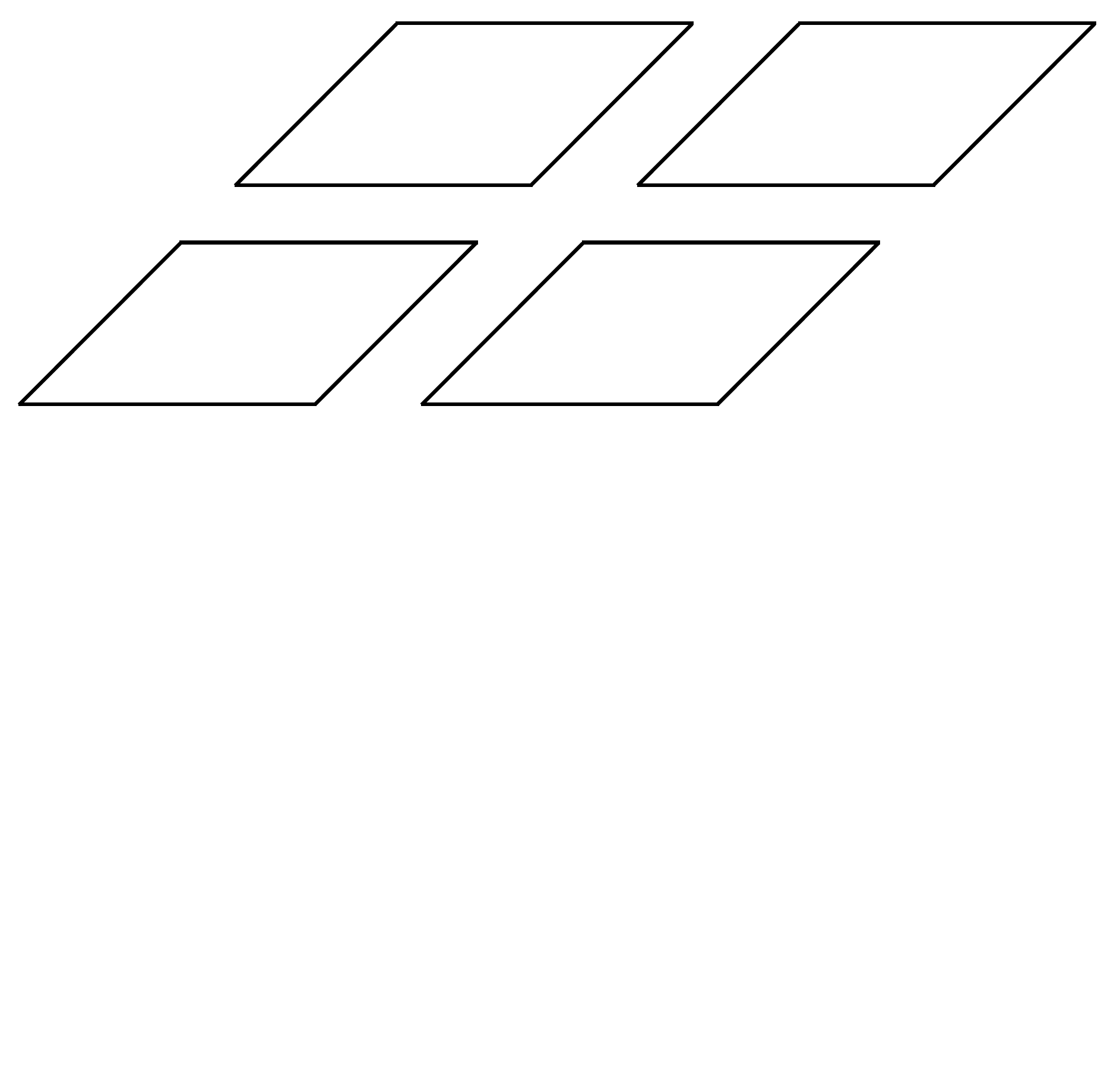}}}
    \hfill
\subfloat[$\mathbf{[d^2;2:2d^2]}$
Using $d^2$ states, 2 observables for $\text{qudit}_1$,
and the usual $2d^2$ observables for $\text{qudit}_2$.
Another PD can be constructed by permuting the two qudits, $\pi{[d^2;2:2d^2]}$.
        \label{b5e}
	]{
	\makebox[0.45\textwidth]{
	\includegraphics[height=1.15in]{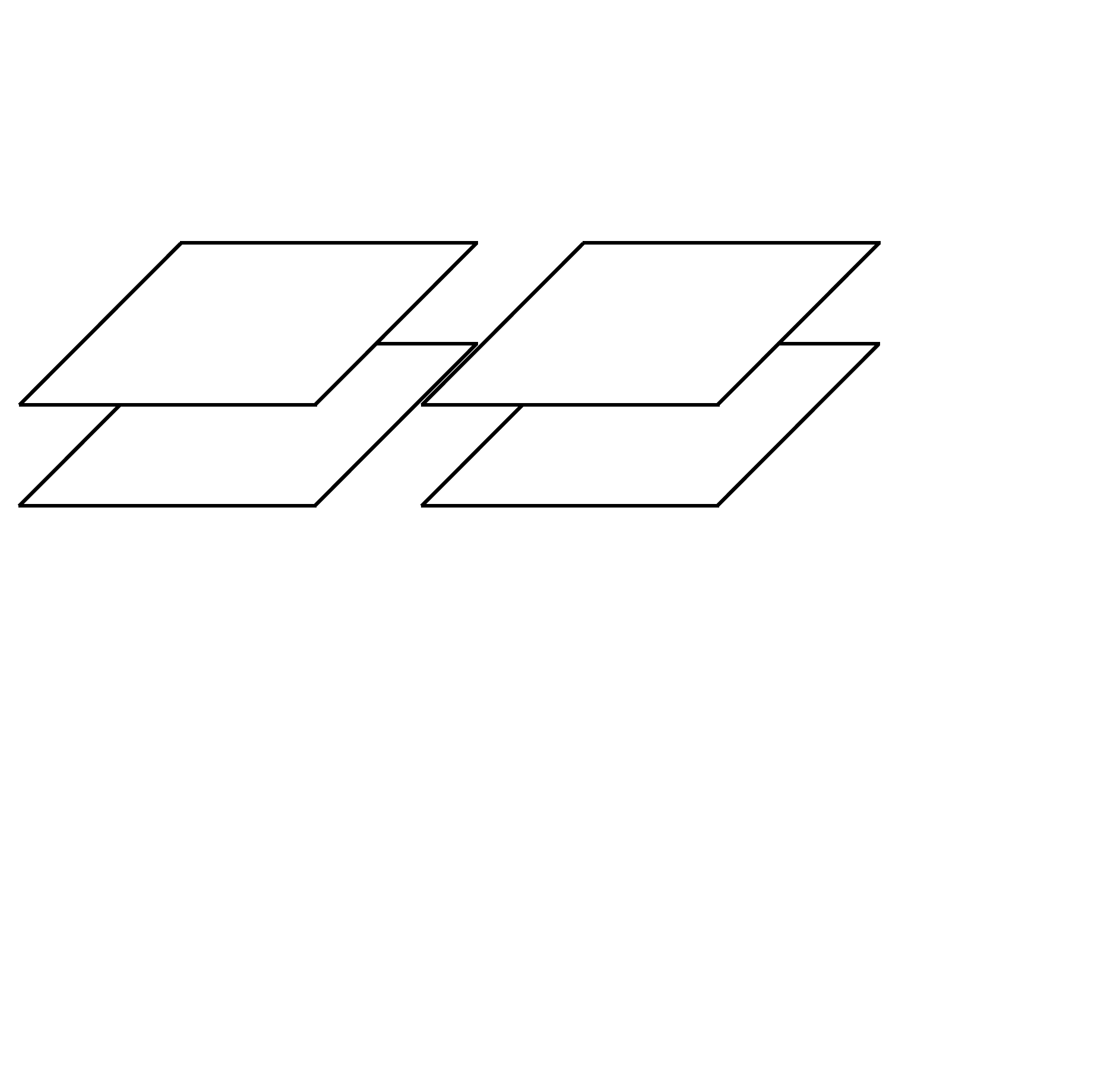}}}
	\\
\subfloat[$\mathbf{[2d;d:2d^2]}$
Using $2d$ states and $d$ $\text{qudit}_1$ observables (such as a $d$-outcome POVM.)
Each stick of butter represents a square matrix that has been rolled up or folded.
        \label{b4e}
	]{
	\makebox[0.45\textwidth]{
	\includegraphics[height=1.15in]{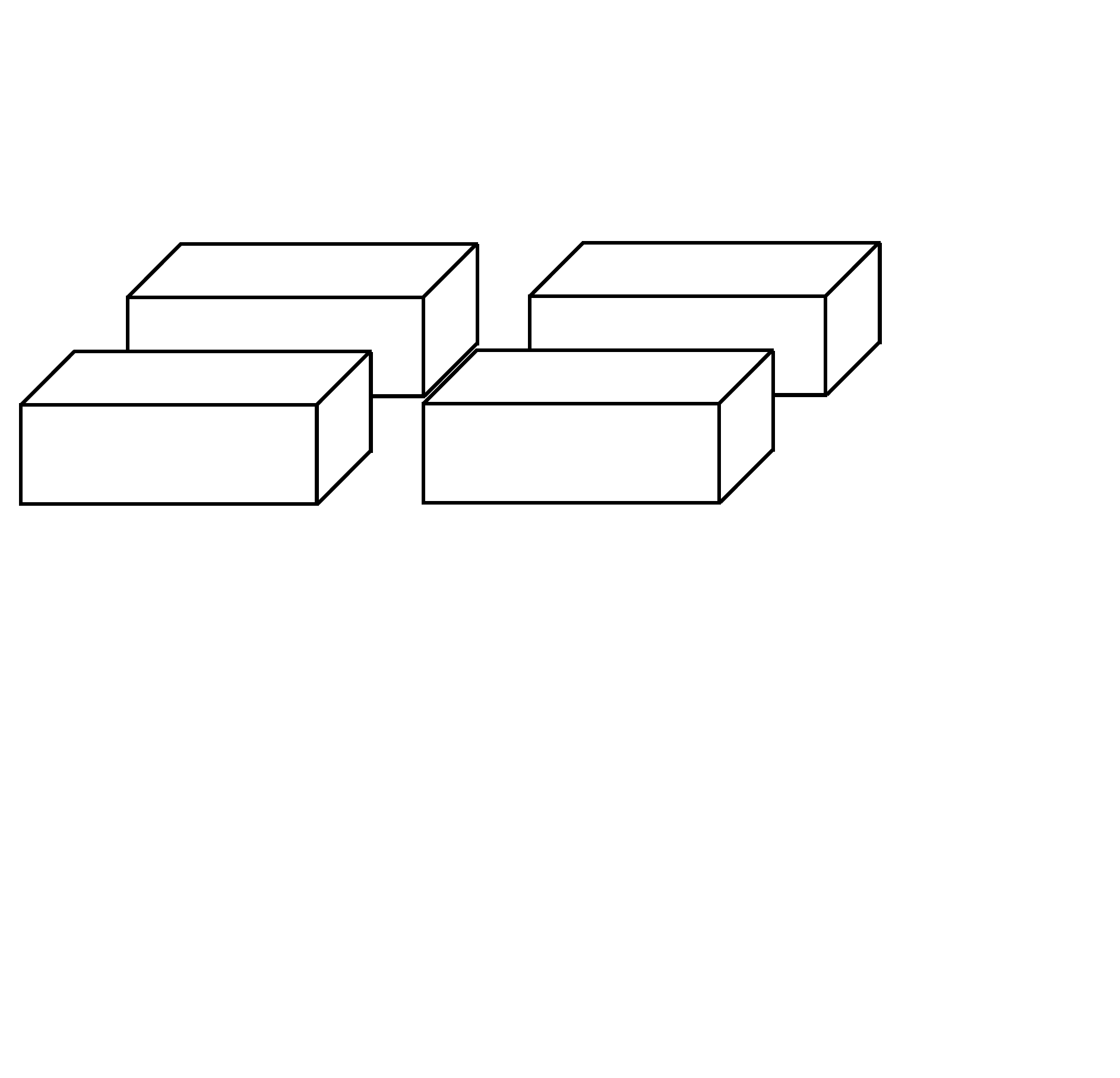}}}
	\hfill
\subfloat[$\mathbf{[d;2d:2d^2]}$
Using $d$ states and $2d$ $\text{qudit}_1$ observables (such as two $d$-outcome POVMs.)
Another PD can be constructed by using vertical sticks of butter, ${\pi[d;2d:2d^2]}$.
        \label{b3e}
	]{
	\makebox[0.45\textwidth]{
	\includegraphics[height=1.15in]{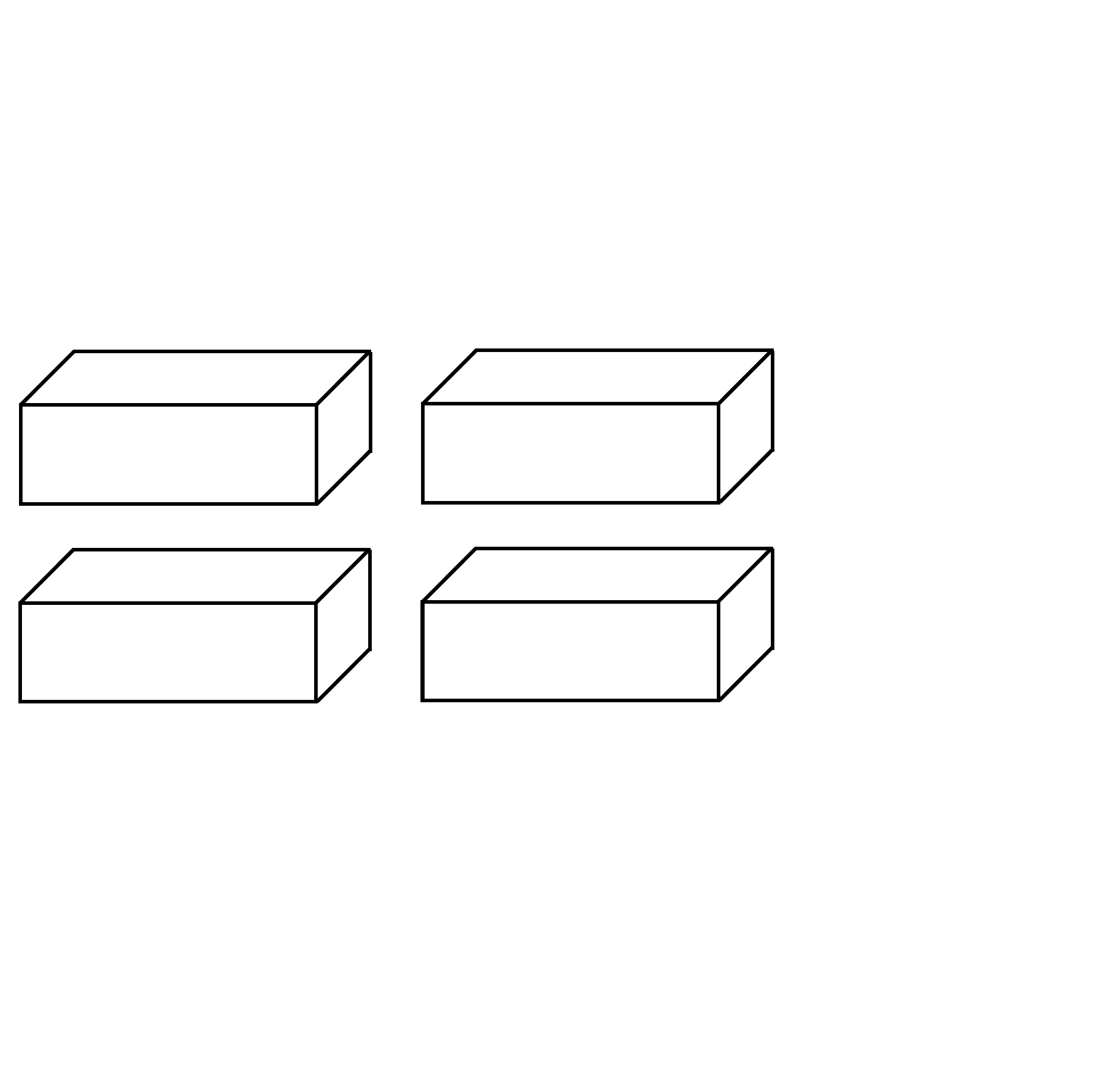}}}
	\\
\subfloat[$\mathbf{[2;d^2:2d^2]}$
Using two states and $d^2$ $\text{qudit}_1$ observables.
Another protocol exists by permuting measurement locations.
        \label{b2e}
	]{
	\makebox[0.45\textwidth]{
	\includegraphics[height=1.15in]{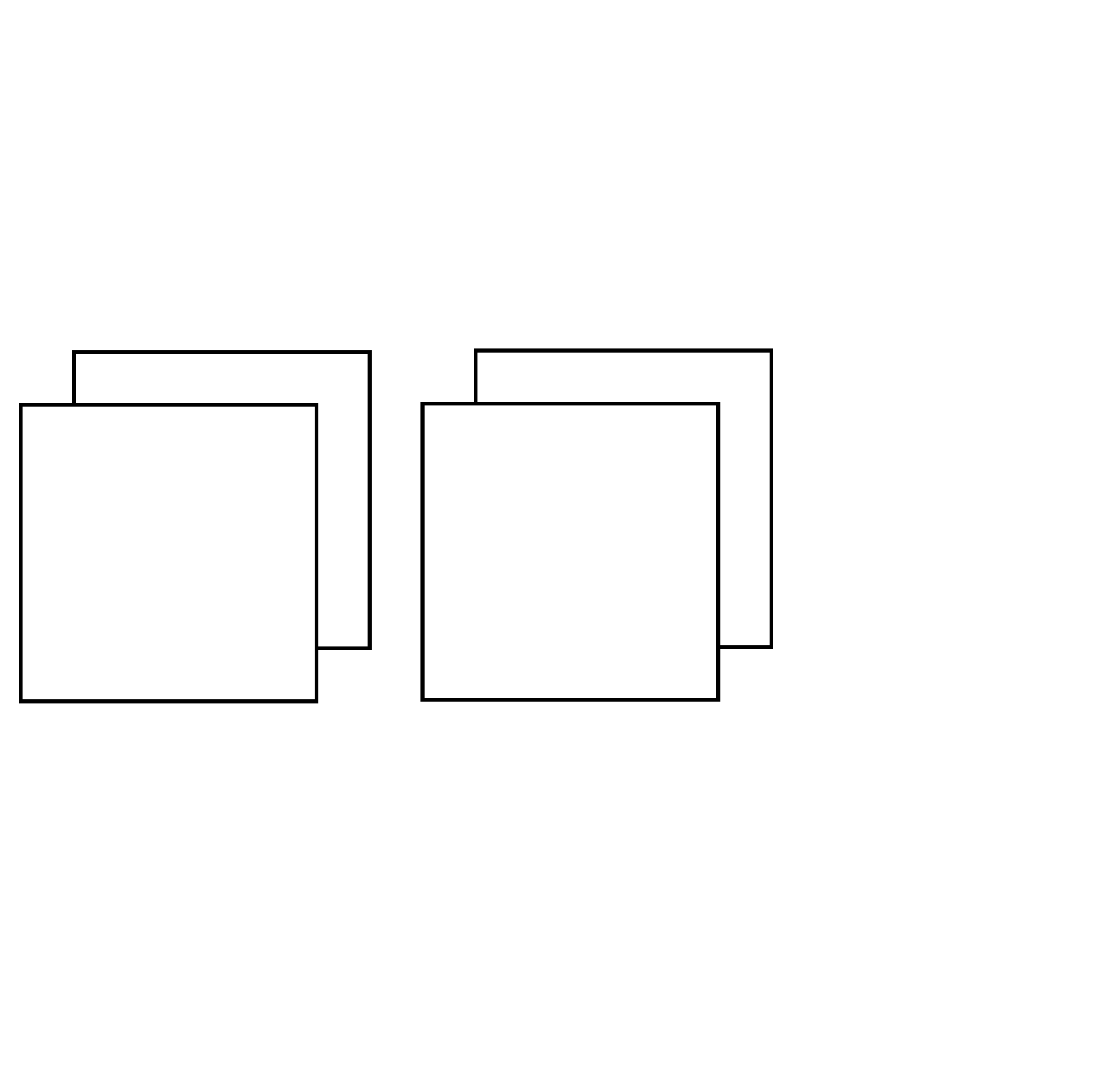}}}
	\hfill
\subfloat[$\mathbf{[1;2d^2:2d^2]}$
Using one state and $2d^2$ and $\text{qudit}_1$ observables .
This particular protocol is symmetric under permuting qudits, $\pi{[1;2d^2:2d^2]=[1;2d^2:2d^2]}$.
        \label{begin}
	]{
	\makebox[0.45\textwidth]{
	\includegraphics[height=1.15in]{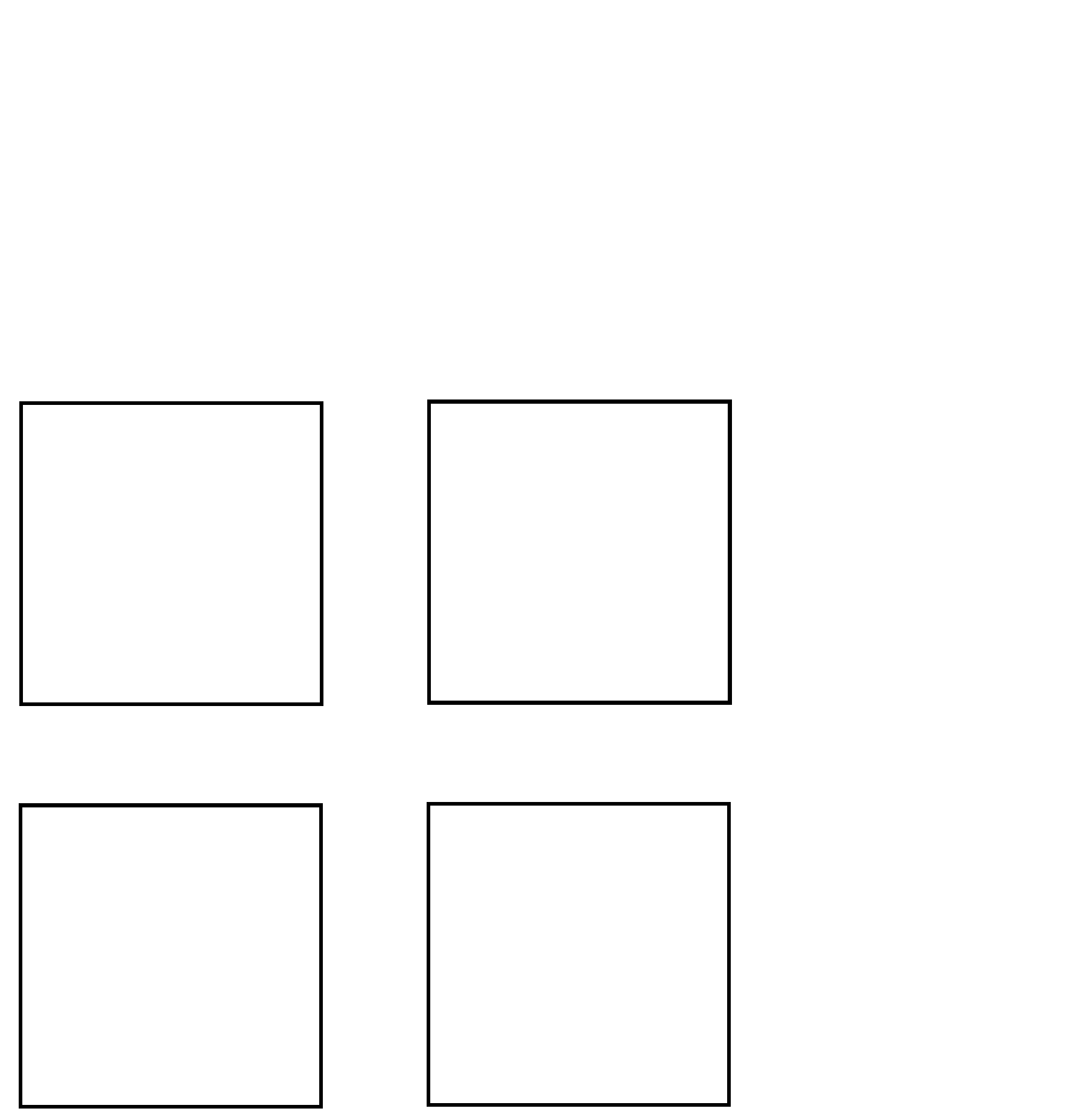}}}
	\\
    	\caption{At the top is a coordinate system for the entries of the data ${S_a}^{ij}$ where $a$ is a state setting, $i$ is a $\text{qudit}_1$ measurement setting, and $j$ is a $\text{qudit}_2$ measurement setting.
	Regions covered by the various shapes represent collected data.
	Each subfigure can be associated with a different measurement protocol one can consider which may further suggest different models of correlation.
    	Each shape corresponds to a corner template while the arrangement of the 4 copies correspond to a square one can translate through.
	The permuted versions of each PD corresponding to the above subfigures are all distinct, except for the PD of subfigure \ref{begin}.
	}
	\label{Wonderful}
\end{figure}

\begin{figure}[h!]
\begin{tabular}{c|cc}
\text{Corners} & \multicolumn{2}{c}{\text{Squares}}\\
	\hline
\subfloat[${[d^2;1:d^2]}$\newline ``Vertical~Plate''  \label{vert}]{
	\makebox[0.3\textwidth]{
	\includegraphics[width=1.1in]{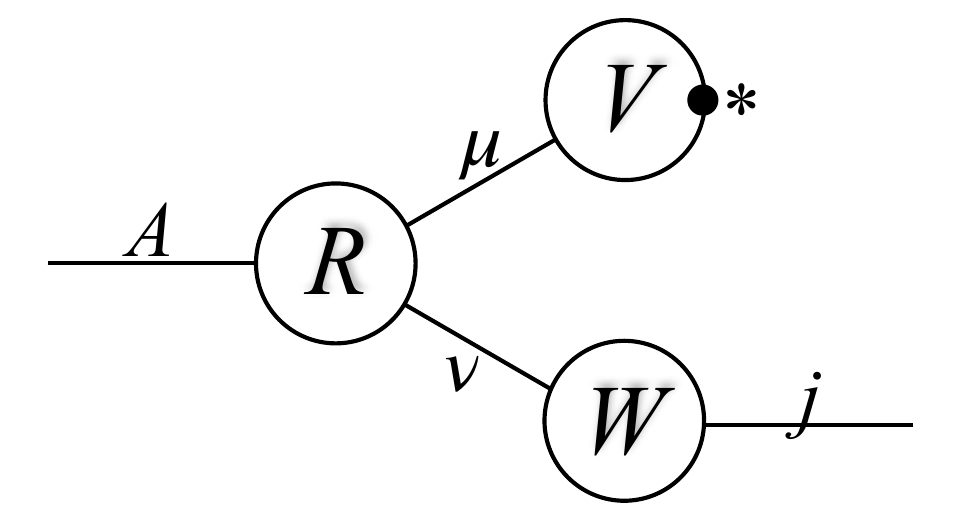}}}&
\subfloat[${[2d^2;1:2d^2]}$\newline ``$di = 0$'']{
	\makebox[0.3\textwidth]{
	\includegraphics[width=1.0in]{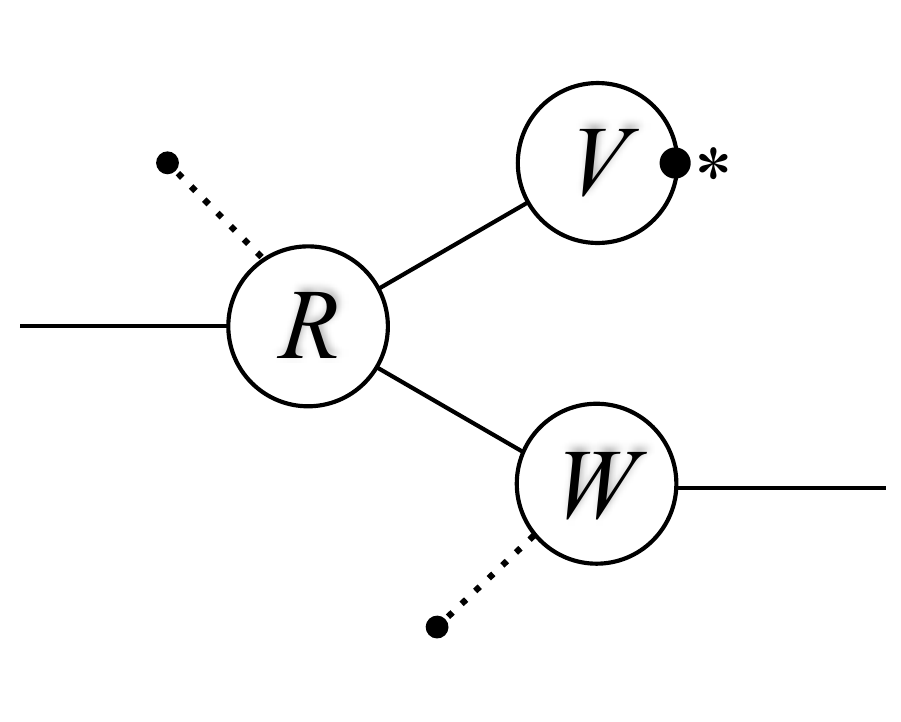}}}&
\subfloat[${[d^2;2:2d^2]}$\newline ``$da = 0$'']{
	\makebox[0.3\textwidth]{
	\includegraphics[width=1.0in]{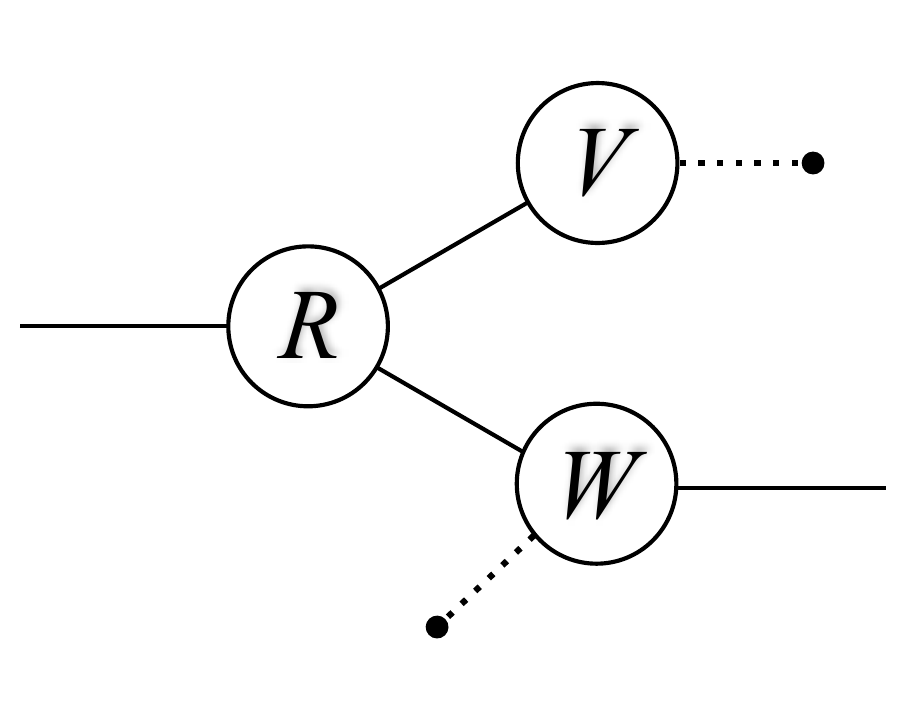}}}\\
\subfloat[${[d;d:d^2]}$\newline ``Stick~of~Butter" \label{butt}]{
	\includegraphics[width=1.00in]{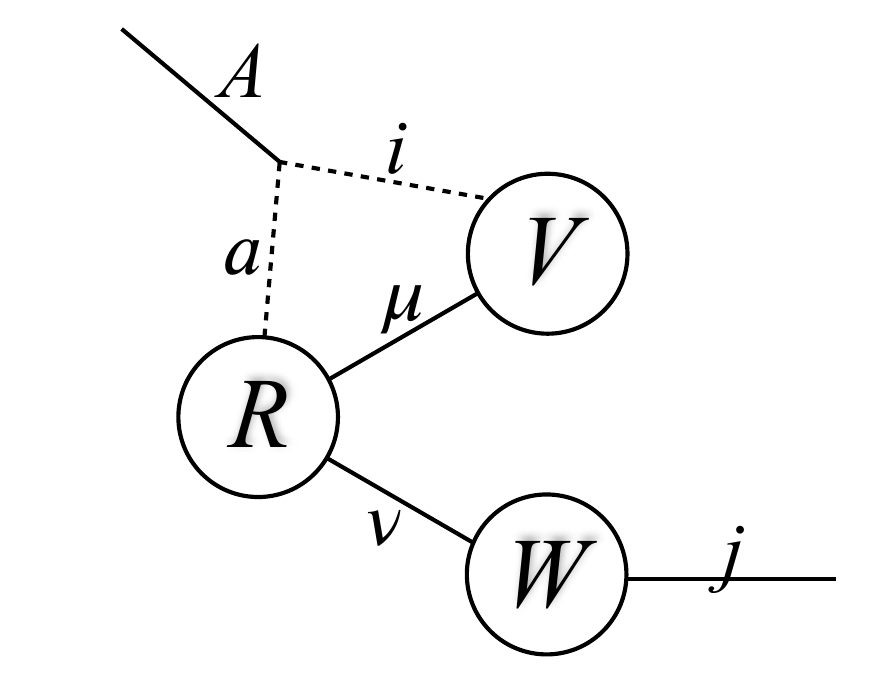}}&
\subfloat[${[2d;d:2d^2]}$\newline ``$di = 0$"]{
	\includegraphics[width=1.0in]{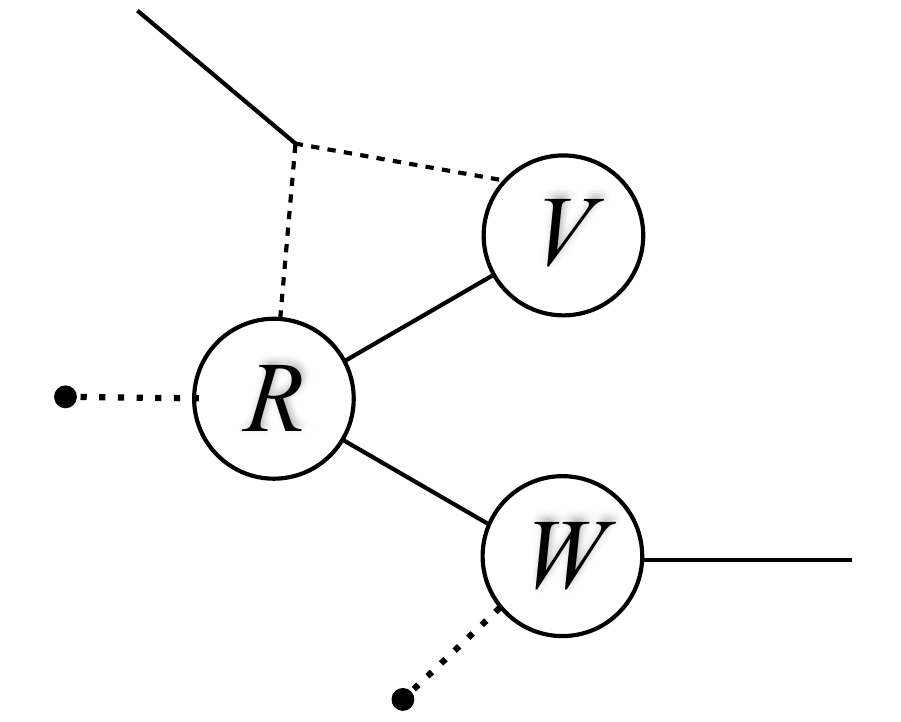}}&
\subfloat[${[d;2d:2d^2]}$\newline ``$da = 0$"]{
	\includegraphics[width=1.0in]{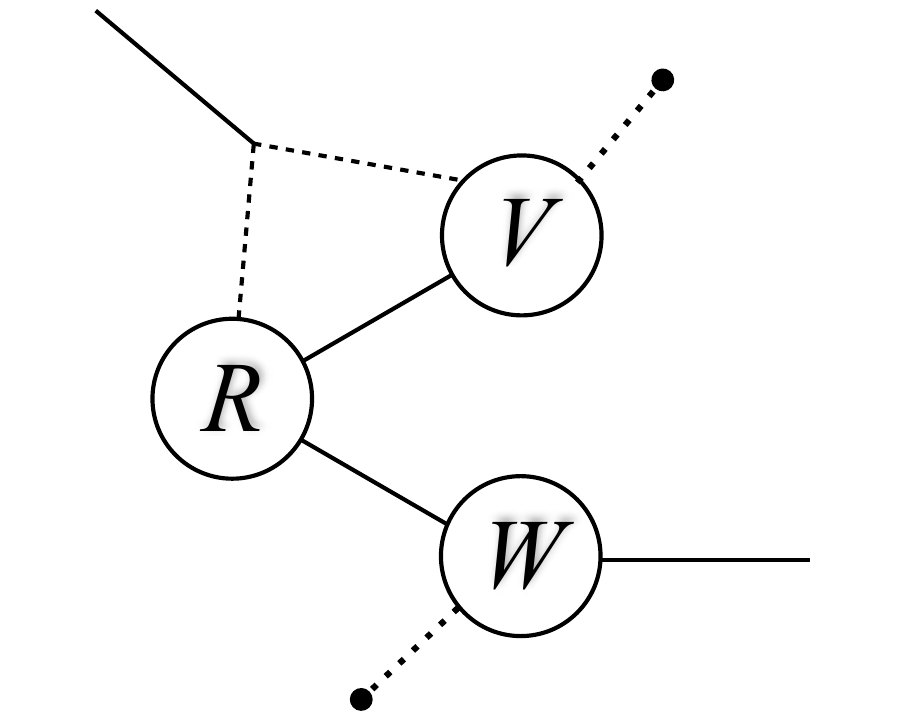}}\\
\subfloat[${[1;d^2:d^2]}$\newline ``Horizontal~Plate" \label{hori}]{
	\includegraphics[width=1.1in]{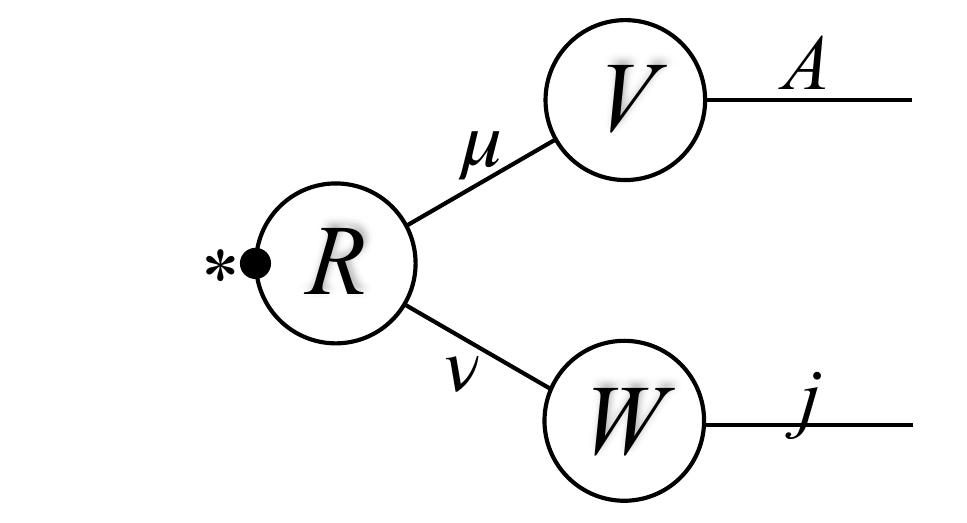}}&
\subfloat[${[2;d^2:2d^2]}$\newline ``$di = 0$'']{
	\includegraphics[width=1.0in]{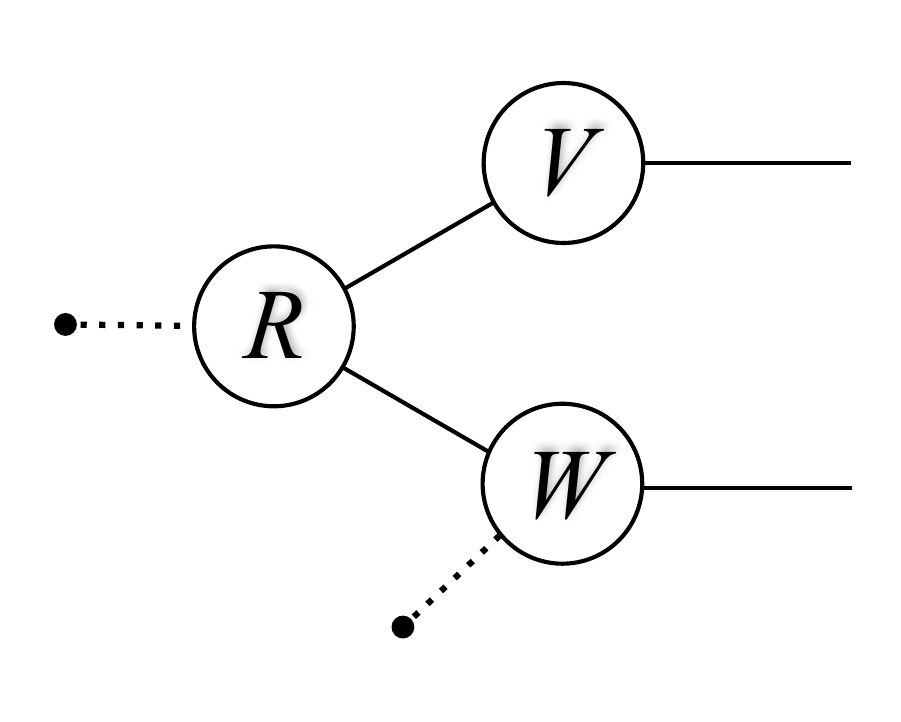}}&
\subfloat[${[1;2d^2:2d^2]}$\newline ``$da = 0$"]{
	\includegraphics[width=1.0in]{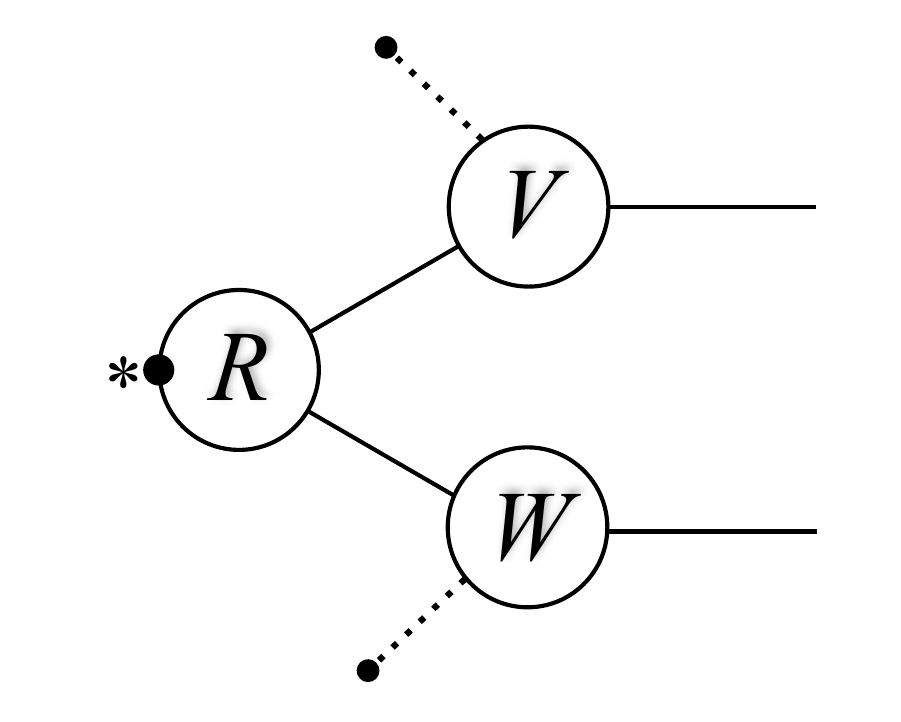}}\\
\end{tabular}
\caption{Diagrammatic representations of PD constructions as arranged in Table \ref{6table}.
Circles represent device parameters.
External lines represent experimental settings.
Internal lines represent a sum over the number of independent model parameters.
The backbone of each diagram, Figure \ref{pic1}, represents the hypothesis that the data ( of an experiment like Figure \ref{2qudev}) can be modeled by Equation \pref{2qu}.
Solid lines represent a range of $d^2$, dashed lines have range $d$, and amputated lines are single valued.
A vertex joining one solid line with two dashed lines represents the delta function, Equation \pref{fancydelta}.
Dotted lines with small circular endpoints correspond to a `$2$', a setting used to displace or distinguish corners.
}
\label{corners}
\end{figure}

Corners and squares can now be made in several ways.
Corners must be $d^2 \times d^2$ (because the data must have rank $\le d^2$ if Equation \pref{snip2} exists.)
There are 3 ways one can do this because we must take $M=d^2$ while there are 3 different ways to make $d^2$
effective states, $[N;L] = [d^2;1]$, $[d;d]$, and $[1;d^2]$, (restricting ourselves to nice multiples.)
Then there are 2 ways each to make a square, $[2N;L :2M]$ or $[N;2L :2M]$, (ignoring that one could mix corner types in a single PD.) (See Table \ref{6table}.)
Having picked one of the 2 qudits, there are almost $2 \times 6=12$ PDs, except that $\pi[1;2d^2 :2d^2] = [1;2d^2 :2d^2]$ is actually a symmetric construction.
So there are $12-1=11$ in total PDs of the type $[N;L:M]$.
To make the construction of these PDs as clear as possible, Figures \ref{corners} and \ref{Wonderful} are given to go over each of them individually.

In Figures \ref{Wonderful}, it is important to recall the distinction between device settings and a device parameters.
Device settings are the external controls that are available in an experiment,
while device parameters are the model dependent numbers used to describe the behavior of the experiment.
Changes in settings can be understood as generating changes in parameters,
but only in a local sense (from corner to corner) which one might not be able to integrate to a global correspondence (because there could be correlations.)
In other words, a constraint such as ``keeping state parameters fixed'' can still be operationally defined but will in general be a non-holonomic constraint.
Similar distinctions are represented mathematically in other physical theories as well, a discussion of which may be found in \cite{nonholo1}.

In Figures \ref{corners}, Corners have been given qualitative names for how they ``fill'' the space of settings as represented in Figures \ref{Wonderful}.
Solid lines represent a range of $d^2$, dashed lines have range $d$, and amputated lines are single valued.
A vertex joining one solid line with two dashed lines represents the delta function
\begin{equation}\label{fancydelta}
\delta_A^{ab} =
	\begin{cases}
	1 &  A = ad+b \\
	0 & \text{otherwise}
	\end{cases}
\end{equation}
where $A\in\{0,1,\ldots,d^2-1\}$ is the solid line and $a,b\in\{0,1,\ldots,d-1\}$ are the dashed lines.
Dotted lines with small circular endpoints represent the settings used to displace the corners of a square, i.e. the `$2$'s in square bracket notation.
Squares have been further labelled based on how they are oriented in the setting dimensions
as represented by the placement of `$2$'s in bracket notation as well as in Figures \ref{Wonderful}.

Further in Figures \ref{corners}, the backbone (Figure \ref{pic1}) of each diagram represents the hypothesis that the data is effectively completely uncorrelated.
However, once a corner is assembled, one can then see from the diagram how this hypothesis may be relaxed
	to weaker types of independence that would still give the PD a trivial value.
Digrammatically, this corresponds to the property that the minimum number of lines one must cut in order to detach the external solid lines
	corresponds exactly to the upper bound in the rank.
Moreover, the displacing lines or `2's can empirically suggest different models of correlation for nontrivial values in the corresponding PD.
For example, a nontrivial value for $[2d^2; 1:2d^2]$ suggests $\text{SPAM}_2$ correlations while $[d^2; 2:2d^2]$ suggests (1,2)-nonlocalities.

To summarize, a square bracket notation has been introduced to represent different PDs one can construct for two-qudit (or qu$d^2$it) systems.
Each PD will have a trivial value, $A^\inv B D^\inv C = 1_{d^2 \times d^2}$, if the system is effectively uncorrelated in that corresponding way.
The types of correlation which violate these PDs should be clear from the topology of their effective backbone (see Figures \ref{pic2} and \ref{pic3}.)
The first is that $[N:M_1,M_2]$ PDs are trivial if $\langle\Tr\rho\,(\Sigma\otimes\Sigma)\rangle = \Tr\langle\rho\,\rangle\langle\Sigma\otimes\Sigma\rangle$
and are thus not sensitive to (12-)nonlocalities.
The second is that $[N;L :M]$ PDs are trivial if $\langle\Tr\rho\,(\Sigma\!\otimes\!\Sigma)\rangle = \Tr\langle\rho\,\Sigma\rangle\!\otimes\!\langle\Sigma\rangle$
so are not sensitive to $\text{SPAM}_1$ correlations.
Similarly $\pi[N;L :M]$ are insensitive to $\text{SPAM}_2$ correlations.

\section{More than Two Qudits}

Increasing the number of qudits, $m>2$, there are many more variations in the kinds of corners and squares we can construct
and so there are many more different types of experiments one can do to detect many more different types of correlation.
One fruitful way of classifying PDs (and the corresponding experiments) is by the matrix \emph{rank} that the corresponding square should have in the absence of correlations.
In particular, for qudit measurements on qud$^m$it there are $m$ types of PDs corresponding to $m$ different ranks, $\rank(M)=d^{2k}$ for $k=1, \ldots, m$.
(See Figures \ref{pic2}, \ref{pic3}, and \ref{pics3} and Tables \ref{3qPD2qish} and \ref{chPD3q}.)
Remember that the rank also determines how the number of experimental settings scales, namely, as ``pairs of settings'' = $\rank(M)^2=d^{4k}$.

Following our previous notation, these classes will be denoted with square brackets by
\begin{equation}
	\Delta_k = [N;L_1, \ldots, L_{m-k}:M_1,\ldots,M_k]
\end{equation}
The generic PD corresponds to $k=m$ which has only 1 corner type (because all the measurement devices are to the right of the colon)
and $m$ square types (because there are m devices to the right of the colon which can be used for displacement) as in section \ref{gen}.
Those PDs which demand the least number of experimental settings correspond to $k=1$ for which there are 4 kinds of corner, 10 kinds of square, and $\frac{1}{2}m(7m^2-12m+7)$ permutational variants, as will be explained.
All of the main variations in $k=1$ are present for $m=3$, so we will start there.
We will also briefly include $k=2$ for $m=3$ qudits to make the construction of the more general PDs clear.

\subsection{Three Qudits}

\begin{figure}[h!]
\subfloat[${[N:\!M_1,M_2,M_3]}$\newline
	Uncorrelated $\leftrightarrow$ rank $\le d^6$]{
	\makebox[0.3\textwidth]{
\includegraphics[height=0.7in]{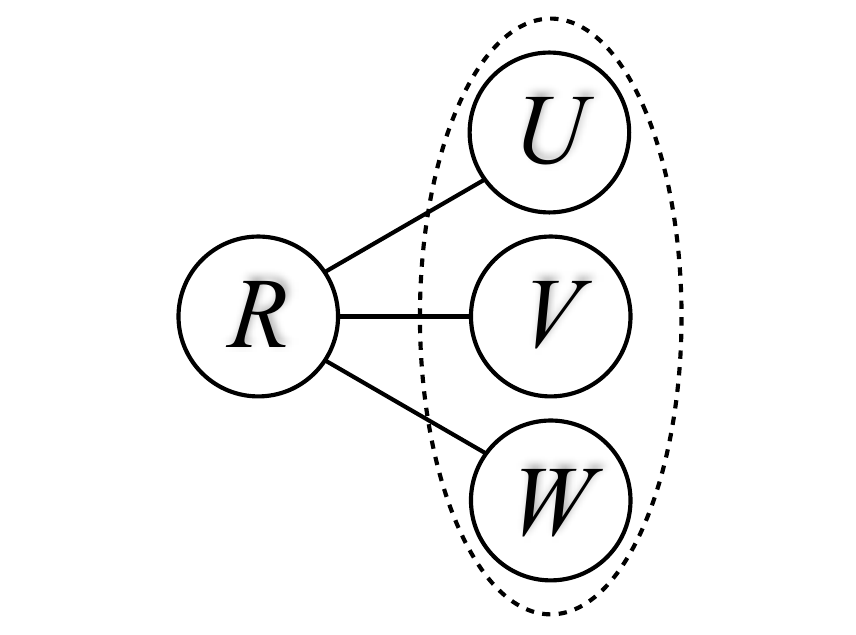}}}
\subfloat[${[N;L:M_1,M_2]}$\newline
	Uncorrelated $\leftrightarrow$ rank $\le d^4$]{
	\makebox[0.3\textwidth]{
\includegraphics[height=0.7in]{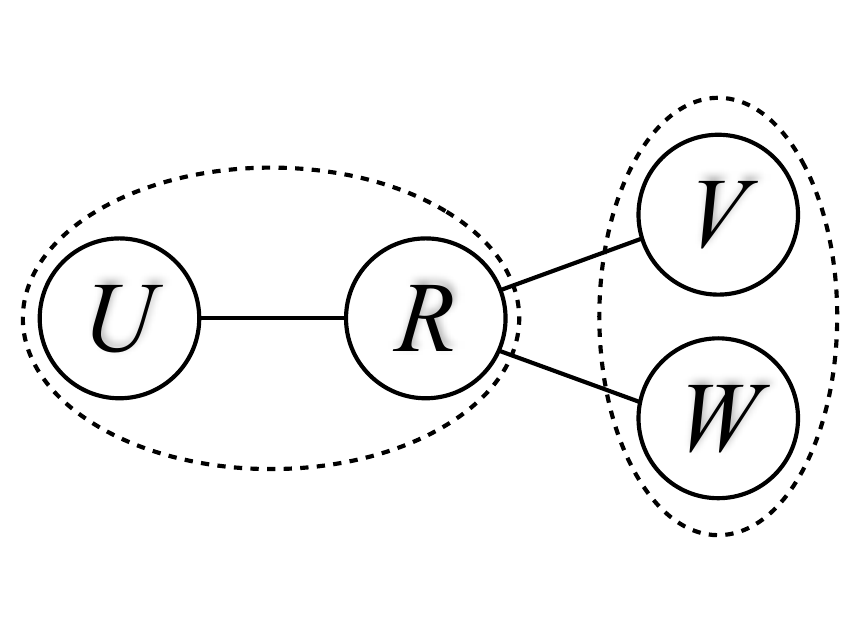}}}
\subfloat[${[N;L_1,L_2 :M]}$\newline
	Uncorrelated $\leftrightarrow$ rank $\le d^2$]{
	\makebox[0.3\textwidth]{
\includegraphics[height=0.7in]{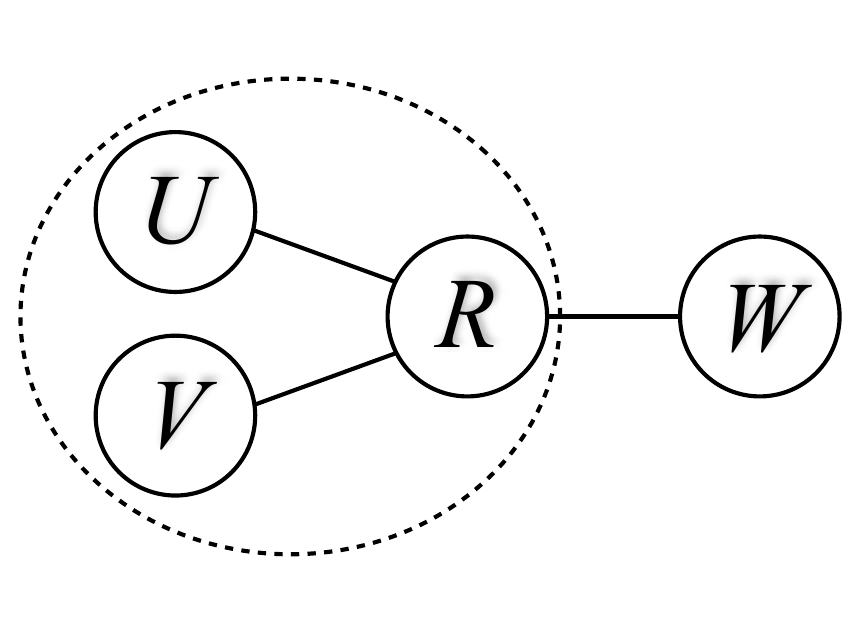}}}
	\caption{Each circular vertex also has an implied external index attached to it like in Figures \ref{pic2} \& \ref{pic3}.
	Each line cut by the dotted separation represents a sum over $d^2$ degrees of freedom.
	These factors determine the upper bound on the rank of the data respectively organized.
	\label{pics3}}
\end{figure}

For three qudits, the data has $1+3=4$ indices or device settings.
If the data is completely uncorrelated, then we may write
\begin{equation}
	{S_a}^{ijk} = R_a^{\lambda\mu\nu}{U_\lambda}^i{V_\mu}^j{W_\nu}^k.
\end{equation}
Such data can be organized into a matrix in 3 basic ways as represented in Figure \ref{pics3}.
These 3 ways further represent separate classes of PD one can construct, each of which are sensitive to different correlations.
Generic PDs, $[N:M,M,M]$, are insensitive to all $(p,q)$-nonlocalities.
The ``$k=2$'' PDs, $[N;L:M,M]$, are insensitive to $(2,3)$-nonlocality and $\text{SPAM}_1$ correlation,
but are sensitive to $(1,2)$-nonlocality, $(1,3)$-nonlocality, $\text{SPAM}_2$ correlation, and $\text{SPAM}_3$ correlation.
The most scalable PDs, $[N;L,L:M]$, are insensitive to $(1,2)$-nonlocalities, $\text{SPAM}_1$, and $\text{SPAM}_2$ correlations.
but are sensitive to $(1,3)$-nonlocality, $(2,3)$-nonlocality, and $\text{SPAM}_3$ correlation.

Of course, one can permute the qudits to make similar statements.
Perhaps the best way to denote each of these is by $\pi[N; \ldots M]$ where now $\pi$ could denote any permutation of 3 elements.
Further, we may denote each $\pi$ the most succinctly with cyclic notation.
For example $(123)[N;L:M,M]$ PDs are insensitive to $31$-nonlocalities and $\text{SPAM}_2$ correlation.
This notion is important for discussing PD symmetries which brings us to the discussion on the ways one can construct corners and squares.

Generic PDs, $[N:M,M,M]$, have no variability in corner types and only 1 basic kind of square, 3 considering which qubit you choose to displace the measurement dimension.
These can be represented in permutation notation as $\Delta$, $(12)\Delta$, and $(13)\Delta$ where $\Delta=[2d^6:2d^2,d^2,d^2]$.
The permutations $\{1, 12, 13\}$ represent the coset for the subgroup $\{1,23\}$ corresponding to the symmetry $(23)\Delta = \Delta$.

\begin{table}[h!]
\vspace{-10pt}
\begin{equation}
\begin{array}{c|cc}
\text{Corners} & \multicolumn{2}{c}{\text{Squares}}\\
{[N;L :M_1,M_2]} & {[2N;L: 2M_1,M_2]} & {[N;2L: 2M_1,M_2]} \\\hline
{[d^4;1:d^2,d^2]} & {[2d^4;1:2d^2,d^2]} & {[d^4;2:2d^2,d^2]}\\
{[d^3;d :d^2,d^2]} & {[2d^3;d :2d^2,d^2]} & {[d^3;2d :2d^2,d^2]}\\
{[d^2;d^2 :d^2,d^2]} & {[2d^2;d^2 :2d^2,d^2]} & {[d^2;2d^2 :2d^2,d^2]}
\end{array}
\end{equation}
\vspace{-10pt}
\caption{$[N;L :M,M]$ PDs require $\mathcal O (d^8)$ settings.
The number of settings is determined by the expected rank of these matrices, $d^4$, for a completely uncorrelated model.
See Figure \ref{pics3}.}\label{3qPD2qish}
\end{table}

For $k=2$ PDs, $[N;L:M,M]$, we have 3 kinds of corner and 6 kinds of square (see Table \ref{3qPD2qish} and compare to Table \ref{6table}.)
We even continue to have the symmetry $(12)\Delta=\Delta$ for $\Delta = [d^2;2d^2 :2d^2,d^2]$.
Except now, a square can be displaced in 3 measurement dimensions.
This gives a total of $3 \times 11 = 33$ partial determinants, 6 per square except for the one with a symmetry (which only only gives $6/2=3$.)
For $m$ qubits, this would be $11\binom{m}{2}$ PDs.

Qudits given a ``1'' in square bracket notation can be considered a trace over that qudit, i.e. choose the identity observable.
Most practical instances will consider qubits, $d=2$, in which case it is important to remember a ``$d$'' and a ``$2$'' are still different in that
a $d$ refers to settings used to make a single kind of corner while a $2$ is used to displace different corners in a square.
Diagrams could be drawn as before to represent corners and squares where we would go on to interpret what trivial values for such PDs mean.

\begin{widetext}

\begin{table}[h!]
\vspace{-10pt}
\begin{equation}
\begin{array}{c|ccc}
\text{Corners} & \multicolumn{3}{c}{\text{Squares}}\\
{[N;L_1,L_2 :M]} & {[2N;L_1,L_2 :2M]} & {[N;2L_1,L_2 :2M]} & {[N;L_1,2L_2 :2M]}\\\hline
{[d^2;1,1 :d^2]} & {[2d^2;1,1 :2d^2]} & {[d^2;2,1 :2d^2]} &  \\
{[d;d,1 :d^2]} & {[2d;d,1 :2d^2]} & {[d;2d,1 :2d^2]} & {[d;d,2 :2d^2]} \\
{[1;d^2,1 :d^2]} & {[2;d^2,1 :2d^2]} & {[1;2d^2,1 :2d^2]^*} & {[1;d^2,2 :2d^2]} \\
{[1;d,d :d^2]} & {[2;d,d :2d^2]^{**}} & {[1;2d,d :2d^2]} &  \\
\end{array}
\end{equation}
\vspace{-10pt}
\caption{$[N;L,L : M]$ PDs require $\mathcal O (d^4)$ settings.
The number of settings is determined by the expected rank of these matrices, $d^2$, for a completely uncorrelated model.
See Figure \ref{pics3}.}\label{chPD3q}
\end{table}

\end{widetext}

\begin{figure}[h!]
\subfloat[${[d^2;1,1:d^2]}$]{
	\makebox[0.24\textwidth]{
\includegraphics[height=0.7in]{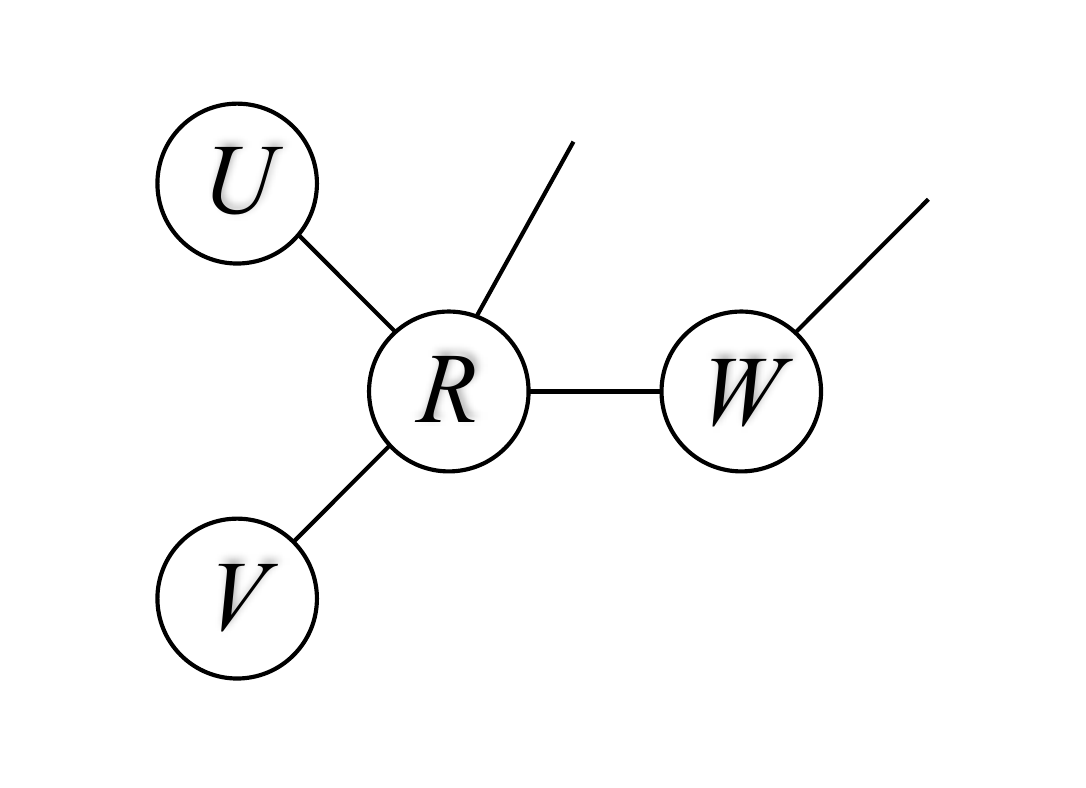}}}
\subfloat[${[d;d,1:d^2]}$]{
	\makebox[0.24\textwidth]{
\includegraphics[height=0.7in]{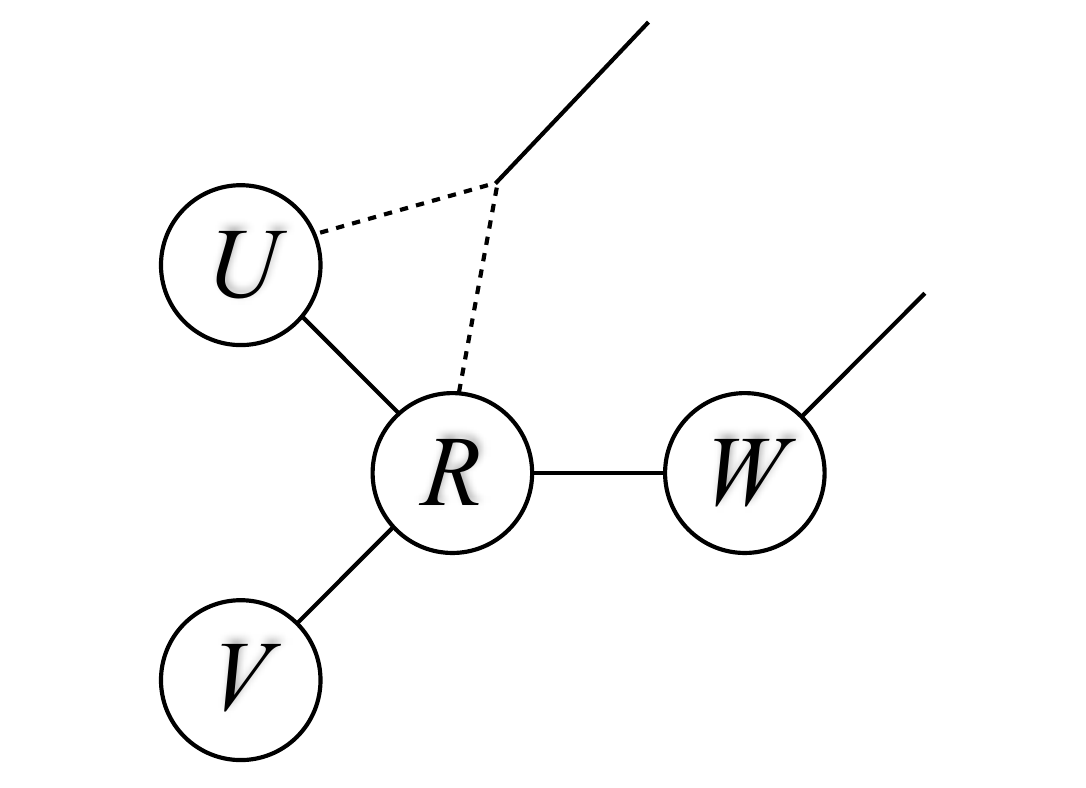}}}
\subfloat[${[1;d^2,1:d^2]}$]{
	\makebox[0.24\textwidth]{
\includegraphics[height=0.7in]{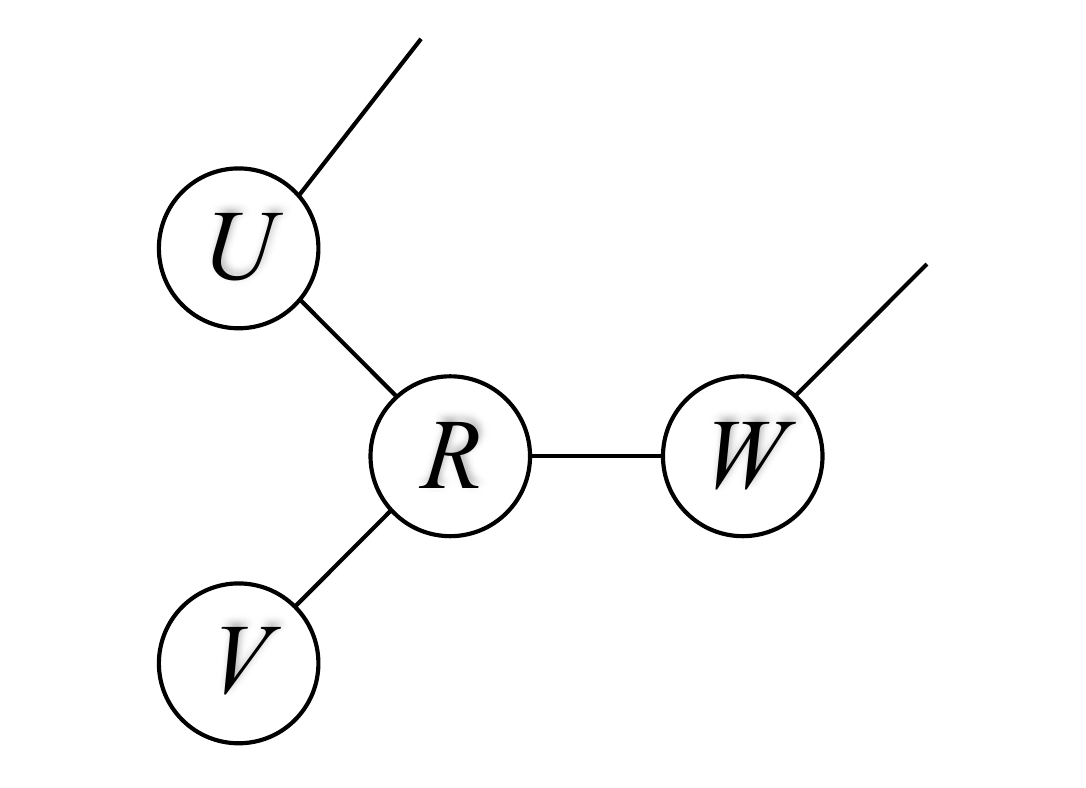}}}
\subfloat[${[1;d,d:d^2]}$]{
	\makebox[0.24\textwidth]{
\includegraphics[height=0.7in]{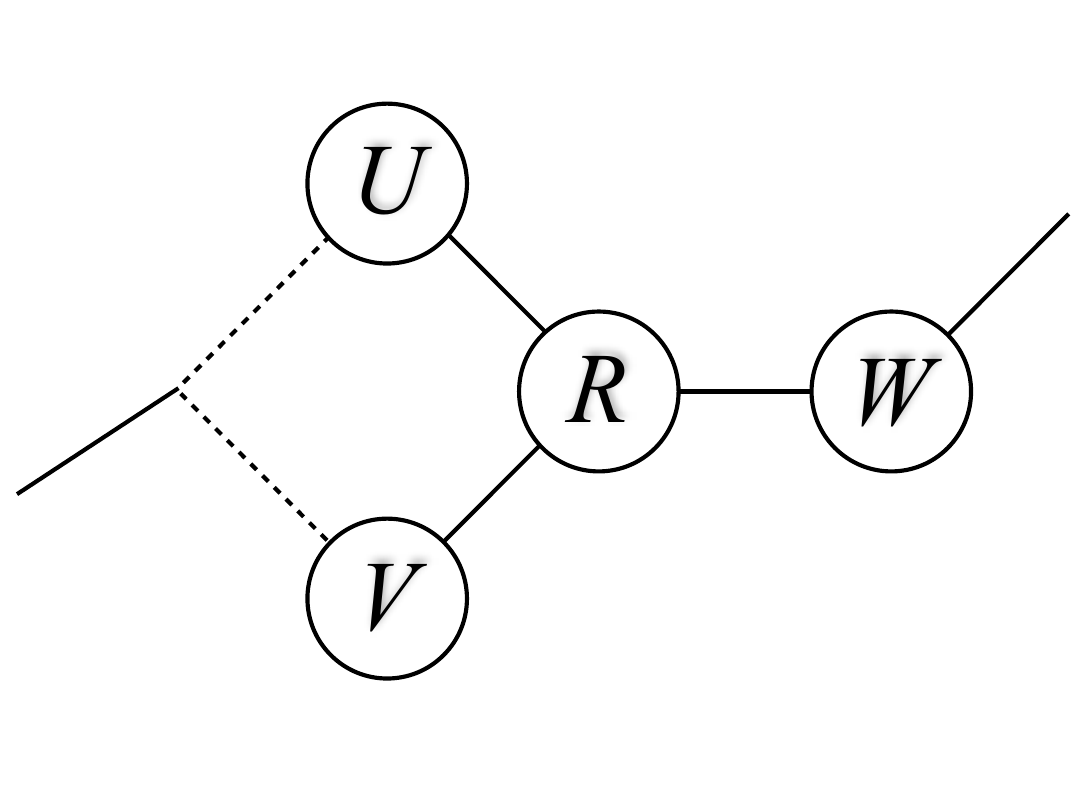}}}
\caption{Diagrams for $[N;L_1,L_2 :M]$ corners, the rows of Table \ref{chPD3q}.
These are the most scalable because the external lines can be detached from each other by cutting a single internal line,
representing the upper bound on the rank of $d^2$ if such a factorization exists.
Further, corners (b) and (d) suggest meaningful PDs of rank $d^3$ because they can factor by cutting $1\frac{1}{2}$ lines.}\label{pics4}
\end{figure}

Finally for the most scalable PDs, $[N;L,L:M]$, we have 4 types of corner and 10 types of square (see Table \ref{chPD3q}.)
Entries kept blank are simply because they are equivalent by permutation with the entry to the left in the table.
Entries marked with an asterisk have a symmetry.
All together, these make 51 PDs which we will explain the combinatorics for in the next section on general $m$.
Illustrating the corners diagrammatically as in Figures \ref{pics4}, we can interpret the meaning of a non-trivial value for each corresponding PD and we will refer to them by there subfigures:
\begin{itemize}
	\item	Rank $d^2$ PDs displaced from $\Delta_a$, $\Delta_b$, and $\Delta_d$ have a trivial value\\ if and only if $\langle RUVW \rangle = \langle RUV \rangle\langle W \rangle$.
	\item	On the other hand, rank $d^2$ PDs displaced from $\Delta_c$ will have a trivial value\\
	if either $\langle RUVW \rangle = \langle RUV \rangle\langle W \rangle$ \emph{or} $\langle RUVW \rangle = \langle U \rangle\langle RVW \rangle$.
	\item	Further, if $\langle RUVW \rangle \neq \langle RUV \rangle\langle W \rangle$ but $\langle RUVW \rangle = \langle U \rangle\langle RVW \rangle$,\\
	then a rank $d^2$ PD displaced from $\Delta_b$ will have a nontrivial value,\\ but one of rank $d^3$ will be trivial.
	\item	Finally, if $\langle RUVW \rangle \neq \langle RUV \rangle\langle W \rangle$ but a PD from $\Delta_d$ but of rank $d^3$ had a trivial value\\
	then either $\langle RUVW \rangle = \langle U \rangle\langle RVW \rangle$ \emph{or} $\langle RUVW \rangle = \langle V \rangle\langle RUW \rangle$
\end{itemize}

\subsection{The Most Scalable $m$ Qudit PDs}

\begin{figure}[h!]
	\centering
\includegraphics[height=0.9in]{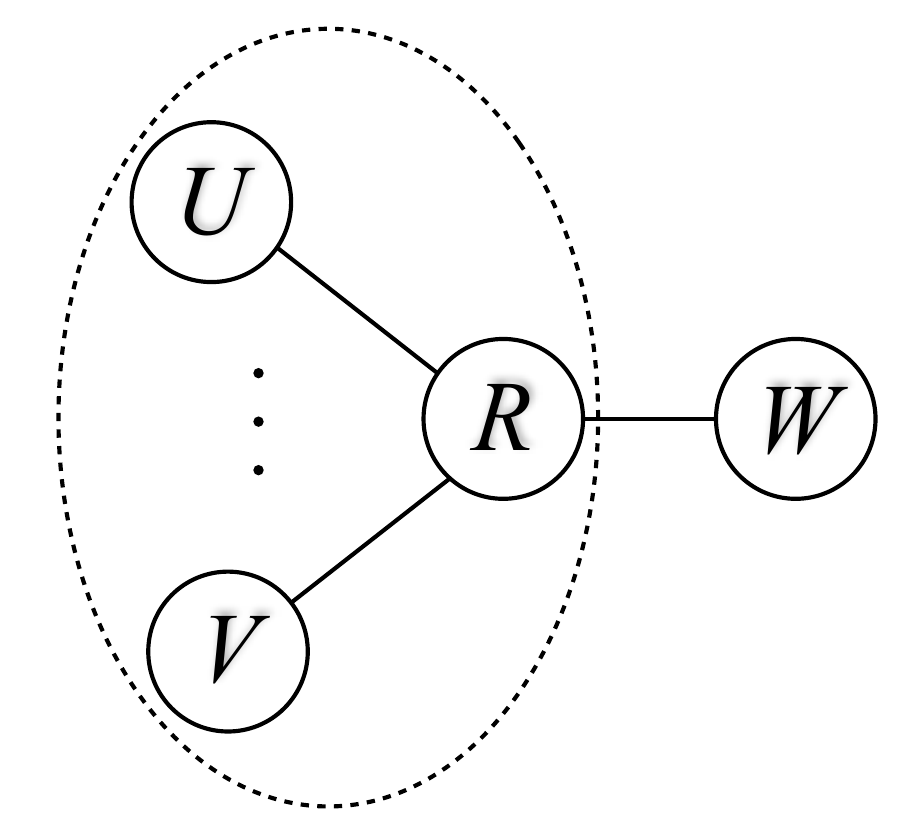}
\caption{There are $\frac{1}{2}m(7m^2-12m+7)$ of the most scalable $m$ qudit PDs, requiring $\mathcal{O}(d^4)$ settings.}\label{wth}
\end{figure}

The PDs of $k=1$ for $m>3$ qudits are essentially no different from $m=3$ because after 3 qudits are chosen, the remaining are fixed to 1 observable or just traced out.
For completeness, let us write the completely uncorrelated data of $m+1$ indices,
\begin{equation}
	{S_a}^{i\ldots jk} = R_a^{\lambda\ldots\mu\nu}{U_\lambda}^i\cdots{V_\mu}^j{W_\nu}^k,
\end{equation}
and represent the $k=1$ class of PDs diagrammatically in Figure \ref{wth}.
Also, we include Tables \ref{chPDmq} which are just like the last section except with a bunch of ellipses to denote `1's for the remaining qudit measurement settings.
We also include the combinatorics for the permuted variations of each PD.
Trivial values are also interpreted just the same as for 3 qudits.

\begin{widetext}

\begin{table}[h!]
\vspace{-10pt}
\begin{equation}
\begin{array}{c|ccc}
\text{Corners} & \multicolumn{3}{c}{\text{Squares}}\\
{[N;L_1,L_2, ... :M]} & {[2N;L_1,L_2, ... :2M]} & {[N;2L_1,L_2, ... :2M]} & {[N;L_1,2L_2, ... :2M]}\\\hline
{[d^2;1,1, ... :d^2]} & {[2d^2;1,1, ... :2d^2]} & {[d^2;2,1, ... :2d^2]} &  \\
{[d;d,1, ... :d^2]} & {[2d;d,1, ... :2d^2]} & {[d;2d,1, ... :2d^2]} & {[d;d,2, ... :2d^2]} \\
{[1;d^2,1, ... :d^2]} & {[2;d^2,1, ... :2d^2]} & {[1;2d^2,1, ... :2d^2]}^* & {[1;d^2,2, ... :2d^2]} \\
{[1;d,d, ... :d^2]} & {[2;d,d, ... :2d^2]}^{**} & {[1;2d,d, ... :2d^2]} & \
\end{array}
\end{equation}
\begin{equation}
\begin{array}{c|ccc}
\text{Corners} & \multicolumn{3}{c}{\text{Squares}}\\
{[N;L_1,L_2, ... :M]} & [2N;L_1,L_2, ... :2M] & [N;2L_1,L_2, ... :2M] & [N;L_1,2L_2, ... :2M]\\\hline
{[d^2;1,1, ... :d^2]} & m & m(m-1) &  \\
{[d;d,1, ... :d^2]} & m(m-1) & m(m-1) & m(m-1)(m-2) \\
{[1;d^2,1, ... :d^2]} & m(m-1) & m(m-1)/2 & m(m-1)(m-2) \\
{[1;d,d, ... :d^2]} & m(m-1)(m-2)/2 & m(m-1)(m-2) \\
\end{array}
\end{equation}
\vspace{-10pt}
\caption{The most scalable PDs, requiring only $\mathcal O (d^4)$ device settings, are just like those for 3 qudits (Table \ref{chPD3q}) except that there are more of them by qudit permutation. The combinatorics for the distinct permutations of each PD are given in the second table, with a total of $\frac{1}{2}m(7m^2-12m+7)$.}\label{chPDmq}
\end{table}

\end{widetext}

\section{Conclusions and Discussion}

In this paper, we considered non-holonomic tomography and its application to multiqudit systems.
Non-holonomic tomography is the use of partial determinants (PDs) to analyze quantum data for detecting various kinds of correlation in SPAM tomography,
where both state and measurement devices have errors.
We demonstrated that there are a multitude of PDs one can consider which are sensitive in different ways to the various correlations that can occur.
Further, we were able to describe these sensitivities based on the topology of the factorization associated with corresponding notions of an effectively uncorrelated system.
For single qudit measurements on a qud$^m$it state, there are $m$ major classes of PD corresponding to matrix rank, $\rank=d^{2k}$ for $k=1, \ldots, m$.
These ranks in turn determine how many device settings are needed ($\mathcal{O}(\rank^2)=\mathcal{O}(d^{4k})$) to experimentally determine the PD.
Finally, we enumerated the class of PDs which require the least number of experimental settings, $k=1$, for any number of qudits.
Figure \ref{summary} is provided as a logical sketch for the technique of non-holonomic tomography.

\begin{widetext}

\begin{figure}[h!]
\begin{equation}
\begin{array}{ccccc}
$(Uncorrelated) Model$	& \xrightarrow{\text{``Topology''}}	& $Rank $ \le r	& \xleftrightarrow{            \hspace{25pt}         }	& 1_{r \times r}	\\
$(Data) Tensor$	& \xrightarrow{\text{Experimental Protocol}}	& $Matrix$	& \xrightarrow{\text{rank } r \text{ PD}}	& \Delta_{r \times r}
\end{array}
\end{equation}
\caption{A summary of the logic in non-holonomic tomography:
The data collected from a quantum experiment is a tensor, with an index associated with each (state preparation or measurement device.
This tensor can be organized as a matrix or ``square'' in various ways and partial determinants can be calculated for these matrices.
Uncorrelated devices correspond to a specific factorization model of the data.
The ``topology'' of this factorization model then sets upper bounds on the rank of any matrix organized from the data.
The rank of these matrices are equal to their upper bound if and only if the partial determinant of their correspondingly sized PD is equal to the identity.}
\label{summary}
\end{figure}

\end{widetext}

PDs have been classified by the types of experimental protocols or ``squares'' one can consider.
However, for each square there are still more PDs corresponding to the order in which the settings are actually put into a matrix.
If one considers data from $2d^2 \times 2d^2$ distinct settings, then for a fixed type of square there are actually $[(2d^2)!]^2$ different PDs by permutation of rows and columns.
On the other hand, these PDs are certainly not distinct quantities.
Some permutations result in PDs which are obviously equivalent, up to familiar transformations,
while others result in more obscure equivalences.
Analyzing these various permutationally equivalent PDs may help to determine which more precisely which settings have correlation.

The first group of permutations which give obviously equivalent PDs correspond to the ways one can traverse the corners of a square | 4 starting points times 2 directions.
If the corners are originally $A$, $B$, $C$, and $D$ as in Equation \pref{square}, then the 8 PDs that result are
\begin{equation}\label{eight}
\begin{array}{cccc}
A^\inv B D^\inv C & B D^\inv C A^\inv & D^\inv C A^\inv B & C A^\inv B D^\inv \\
C^\inv D B^\inv A & D B^\inv A C^\inv & B^\inv A C^\inv D & A C^\inv D B^\inv. \\
\end{array}
\end{equation}
PDs in the same row are related by cyclic permutation while those in the same column are inverses of each other.
Each of these PDs are equivalent to each other up to inverse and conjugation | e.g. $(B^\inv A C^\inv D) = (B^\inv A)\,(A^\inv B D^\inv C)^\inv (B^\inv A)^\inv$.
The second group corresponds to those permutations that keep settings within their respective corners:
\begin{widetext}

\begin{equation}
\left[
\begin{array}{cc}
	A & B\\
	C & D
\end{array}
\right]
\longrightarrow
\left[
\begin{array}{cc}
	\pi_{SP1} & 0\\
	0 & \pi_{SP2}\end{array}
\right]
\left[
\begin{array}{cc}
	A & B\\
	C & D
\end{array}
\right]
\left[
\begin{array}{cc}
	\pi_{M1} & 0\\
	0 & \pi_{M2}
\end{array}
\right]^\inv
\end{equation}
where all the $\pi$s are $d^2 \times d^2$ permutation matrices.
There are $(d^2!)^4$ such elements.
These PDs are equivalent to each other up to conjugation since
\begin{equation}
\Delta\left(
\left[
\begin{array}{cc}
	\pi_{SP1} & 0\\
	0 & \pi_{SP2}\end{array}
\right]
\left[
\begin{array}{cc}
	A & B\\
	C & D
\end{array}
\right]
\left[
\begin{array}{cc}
	\pi_{M1} & 0\\
	0 & \pi_{M2}
\end{array}
\right]^\T
\right)
=
\pi_{M1}^\inv
\Delta\left(
\left[
\begin{array}{cc}
	A & B\\
	C & D
\end{array}
\right]
\right)
\pi_{M1}
\end{equation}

\end{widetext}
where $\Delta$ is the standard PD defined by Equations \pref{square} and \pref{PD}.\footnote{
If one considers data from $(d^2+1) \times (d^2+1)$ settings (as described in Appendix \ref{plus1},) then there are $\binom{d^2+1}{2}^2$ distinct PDs,
having already divided out the aforementioned equivalences.
This is because one must choose the $d^2-1$ rows and $d^2-1$ columns of the data that will be common to each corner.
The remaining 2 rows and 2 columns are what actually displace the corners.}
Permutations beyond these two groups ``delocalize'' settings across corners (experiments) and thus give PDs which are equivalent but in a much less obvious way.

Another important comment is that the links between corners, as considered in this paper, have no immediate sense of distance.
This is a consequence of the gauge degrees of freedom.
We don't a priori have the ability to say how different states in experiment A are from states in experiment B, even if they share the same state settings.
However, a notion of distance can be introduced if the devices are also equipped with \emph{continuous} settings.
A discussion of this technique may be found in \cite{nonholo1}, Sections III.A, III.C, and IV.

\begin{figure}[h!]
\centering
	\centering
\includegraphics[width=1.25in]{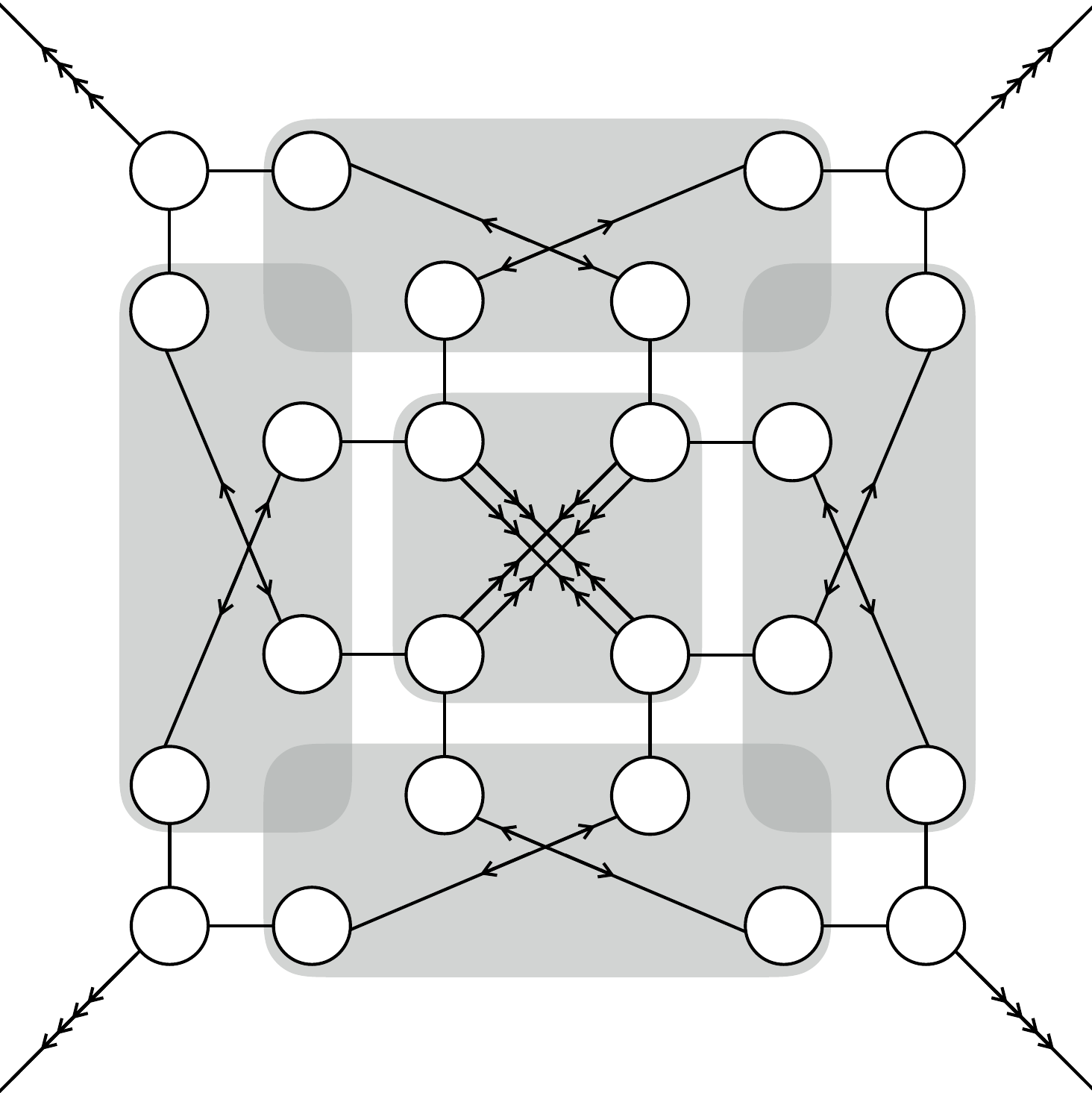}
\caption{A PD for a two qubit system that has the topology of a 2-dimensional surface.
Eight copies of Figure \ref{pic1} can be found along with 4-lined vertices corresponding to contractions with antisymmetric symbols,
$\varepsilon_{\mu\nu\rho\sigma}$, which appear in expressions for matrix inverses.}\label{3shells}
\end{figure}

From a mathematical perspective, it is intriguing that there is this relationship between matrix rank and holonomy.
These holonomies can in fact be generalized to higher-dimensional quantities (like surfaces etc., rather than just loops) which can test for more general tensor ranks.
Such tests can be interpreted as measures of higher $n$-point correlations between devices.
Their construction is relatively simple and requires just one observation:
that matrix inverses just consist of several contractions with antisymmetric tensors (or Levi-Civita epsilon symbols.)
The first author expects to soon publish full details on the method for constructing such quantities (see Figure \ref{3shells}.)

\begin{acknowledgments}
S.J.v.E. was supported in part by ARO/LPS under Contract No. W911NF-14-C-0048.
\end{acknowledgments}

\appendix

\section{An $\mathbf{r \times r}$ PD for an $\mathbf{(r+1)\times(r+1)}$ Matrix}\label{plus1}

In this section, we show how to use PDs to test if an $(r+1)\times(r+1)$ matrix has rank $r$.
Of course, this test should be equivalent to checking if a regular determinant is zero.
However, this construction also generalizes to check if any $s \times s$ matrix has rank $r < s$.
These tests can be associated with experimental protocols which require fewer device settings than the $2r \times 2r$ protocol.
In fact, the $(r+1)\times(r+1)$ protocol has already been applied.\cite{mccormick2017experimental}

Suppose we have an $(r+1)\times(r+1)$ matrix, $S$, which we suspect has rank $\le r$.
We can calculate an $r\times r$ PD by generating a $2r\times2r$ matrix, $\tilde S$, partitioning $S$ as follows
\begin{widetext}

\[
	S = 
	\left[
	\begin{array}{c|ccc|c}
		* & * & * & * & * \\
		\hline
		* & * & * & * & * \\
		* & * & * & * & * \\
		* & * & * & * & * \\
		\hline
		* & * & * & * & * \\
	\end{array}
	\right]
	=
	\left[
	\begin{array}{ccc}
		a & \vec \beta^\T & b \\
		\vec\alpha & \mathbf{M} & \vec \delta \\
		c & \vec \gamma^\T & d \\
	\end{array}
	\right]
	\longrightarrow
	\tilde S
	=
	\left[
	\begin{array}{cc|cc}
		a & \vec \beta^\T & \vec \beta^\T& b \\
		\vec\alpha & \mathbf{M} & \mathbf{M} & \vec \delta \\
		\hline
		\vec\alpha & \mathbf{M} & \mathbf{M} & \vec \delta \\
		c & \vec \gamma^\T & \vec \gamma^\T & d \\
	\end{array}
	\right]
	\equiv
	\left[
	\begin{array}{cc}
		A & B \\
		C & D
	\end{array}
	\right].
\]

\end{widetext}
It is useful to define the following matrices:

\[
\begin{array}{ccc}
	\tilde\alpha
	=
	\left[
	\begin{array}{cc}
		1 & \vec 0^\T \\
		-\mathbf{M}^\inv\vec \alpha & \mathbf{1} \\
	\end{array}
	\right]
	&\hspace{20pt}&
	\tilde\beta
	=
	\left[
	\begin{array}{cc}
		1 & -\vec \beta^\T\mathbf{M}^\inv \\
		\vec 0 & \mathbf{1} \\
	\end{array}
	\right]
	\\
	\vspace{10pt}
	\\
	\tilde\gamma
	=
	\left[
	\begin{array}{cc}
		\mathbf{1} & \vec 0 \\
		-\vec \gamma^\T\mathbf{M}^\inv & 1 \\
	\end{array}
	\right]
	&\hspace{20pt}&
	\tilde\delta
	=
	\left[
	\begin{array}{cc}
		\mathbf{1} & -\mathbf{M}^\inv\vec \delta \\
		\vec 0^\T & 1 \\
	\end{array}
	\right]
\end{array}
\]
(which one may note are representations of $(r-1)$-dimensional translation groups.)
These matrices allow us to partially diagonalize each corner:
\[
\begin{array}{ccc}
	\tilde\beta A \tilde \alpha = 
	\left[
	\begin{array}{cc}
		A/M & \vec 0^\T \\
		\vec0 & \mathbf{M} \\
	\end{array}
	\right]
	&\hspace{20pt}&
	\tilde\beta B \tilde \delta = 
	\left[
	\begin{array}{cc}
		\vec 0^\T& B/M \\
		\mathbf{M} & \vec 0 \\
	\end{array}
	\right]
	\\
	\vspace{10pt}
	\\
	\tilde\gamma C \tilde \alpha = 
	\left[
	\begin{array}{cc}
		\vec0 & \mathbf{M} \\
		C/M & \vec 0^\T \\
	\end{array}
	\right]
	&\hspace{20pt}&
	\tilde\gamma D \tilde \delta = 
	\left[
	\begin{array}{cc}
		\mathbf{M} & \vec 0 \\
		\vec 0^\T & D/M \\
	\end{array}
	\right]
\end{array}
\]
where we denote the Schur complements by
\[
\begin{array}{ccc}
	A/M = a - \vec\beta^T M^\inv\vec\alpha
	&\hspace{10pt}&
	B/M = b - \vec\beta^T M^\inv\vec\delta
	\\
	C/M = c - \vec\gamma^T M^\inv\vec\alpha
	&\hspace{10pt}&
	D/M = d - \vec\gamma^T M^\inv\vec\delta.
\end{array}
\]
All of the various partial determinants can thus be simplified:
\begin{widetext}

\begin{align*}
	A^\inv B D^\inv C &= 1+(x-1)\tilde\alpha & C^\inv D B^\inv A &= \frac{1}{x^2}\Big[1+(x-1)\tilde\alpha^\inv\Big]\\
	B D^\inv C A^\inv  & = 1+(x-1)\tilde\beta^\inv & A C^\inv D B^\inv &= \frac{1}{x^2}\Big[1+(x-1)\tilde\beta\Big]\\
	D^\inv C A^\inv B &= 1+(x-1)\tilde\delta & B^\inv A C^\inv D &= \frac{1}{x^2}\Big[1+(x-1)\tilde\delta^\inv\Big]\\
	C A^\inv B D^\inv & = 1+(x-1)\tilde\gamma^\inv & D B^\inv A C^\inv &= \frac{1}{x^2}\Big[1+(x-1)\tilde\gamma\Big]\\
\end{align*}

\end{widetext}
where
\[
	x = \frac{(B/M)(C/M)}{(A/M)(D/M)}
	= \frac{\det B \det C}{\det A \det D}.
\]\\
One can see that each of these PDs is equal to the identity if and only if $x=1$.
This condition on $x$ must be equivalent to $\Det S = 0$ (given the existence of $M^\inv$.)

\bibliography{Non-Holo_Multi}

%merlin.mbs apsrev4-1.bst 2010-07-25 4.21a (PWD, AO, DPC) hacked
%Control: key (0)
%Control: author (8) initials jnrlst
%Control: editor formatted (1) identically to author
%Control: production of article title (-1) disabled
%Control: page (0) single
%Control: year (1) truncated
%Control: production of eprint (0) enabled
\begin{thebibliography}{19}%
\makeatletter
\providecommand \@ifxundefined [1]{%
 \@ifx{#1\undefined}
}%
\providecommand \@ifnum [1]{%
 \ifnum #1\expandafter \@firstoftwo
 \else \expandafter \@secondoftwo
 \fi
}%
\providecommand \@ifx [1]{%
 \ifx #1\expandafter \@firstoftwo
 \else \expandafter \@secondoftwo
 \fi
}%
\providecommand \natexlab [1]{#1}%
\providecommand \enquote  [1]{``#1''}%
\providecommand \bibnamefont  [1]{#1}%
\providecommand \bibfnamefont [1]{#1}%
\providecommand \citenamefont [1]{#1}%
\providecommand \href@noop [0]{\@secondoftwo}%
\providecommand \href [0]{\begingroup \@sanitize@url \@href}%
\providecommand \@href[1]{\@@startlink{#1}\@@href}%
\providecommand \@@href[1]{\endgroup#1\@@endlink}%
\providecommand \@sanitize@url [0]{\catcode `\\12\catcode `\$12\catcode
  `\&12\catcode `\#12\catcode `\^12\catcode `\_12\catcode `\%12\relax}%
\providecommand \@@startlink[1]{}%
\providecommand \@@endlink[0]{}%
\providecommand \url  [0]{\begingroup\@sanitize@url \@url }%
\providecommand \@url [1]{\endgroup\@href {#1}{\urlprefix }}%
\providecommand \urlprefix  [0]{URL }%
\providecommand \Eprint [0]{\href }%
\providecommand \doibase [0]{http://dx.doi.org/}%
\providecommand \selectlanguage [0]{\@gobble}%
\providecommand \bibinfo  [0]{\@secondoftwo}%
\providecommand \bibfield  [0]{\@secondoftwo}%
\providecommand \translation [1]{[#1]}%
\providecommand \BibitemOpen [0]{}%
\providecommand \bibitemStop [0]{}%
\providecommand \bibitemNoStop [0]{.\EOS\space}%
\providecommand \EOS [0]{\spacefactor3000\relax}%
\providecommand \BibitemShut  [1]{\csname bibitem#1\endcsname}%
\let\auto@bib@innerbib\@empty
%</preamble>
\bibitem [{\citenamefont {Jackson}\ and\ \citenamefont {van
  Enk}(2015)}]{jackson2015detecting}%
  \BibitemOpen
  \bibfield  {author} {\bibinfo {author} {\bibfnamefont {C.}~\bibnamefont
  {Jackson}}\ and\ \bibinfo {author} {\bibfnamefont {S.~J.}\ \bibnamefont {van
  Enk}},\ }\href@noop {} {\bibfield  {journal} {\bibinfo  {journal} {Physical
  Review A}\ }\textbf {\bibinfo {volume} {92}},\ \bibinfo {pages} {042312}
  (\bibinfo {year} {2015})}\BibitemShut {NoStop}%
\bibitem [{\citenamefont {Jackson}\ and\ \citenamefont {van
  Enk}(2017)}]{nonholo1}%
  \BibitemOpen
  \bibfield  {author} {\bibinfo {author} {\bibfnamefont {C.}~\bibnamefont
  {Jackson}}\ and\ \bibinfo {author} {\bibfnamefont {S.}~\bibnamefont {van
  Enk}},\ }\href@noop {} {\bibfield  {journal} {\bibinfo  {journal} {arXiv
  preprint arXiv:1702.00118}\ } (\bibinfo {year} {2017})}\BibitemShut {NoStop}%
\bibitem [{\citenamefont {Merkel}\ \emph {et~al.}(2013)\citenamefont {Merkel},
  \citenamefont {Gambetta}, \citenamefont {Smolin}, \citenamefont {Poletto},
  \citenamefont {C{\'o}rcoles}, \citenamefont {Johnson}, \citenamefont {Ryan},\
  and\ \citenamefont {Steffen}}]{merkel}%
  \BibitemOpen
  \bibfield  {author} {\bibinfo {author} {\bibfnamefont {S.~T.}\ \bibnamefont
  {Merkel}}, \bibinfo {author} {\bibfnamefont {J.~M.}\ \bibnamefont
  {Gambetta}}, \bibinfo {author} {\bibfnamefont {J.~A.}\ \bibnamefont
  {Smolin}}, \bibinfo {author} {\bibfnamefont {S.}~\bibnamefont {Poletto}},
  \bibinfo {author} {\bibfnamefont {A.~D.}\ \bibnamefont {C{\'o}rcoles}},
  \bibinfo {author} {\bibfnamefont {B.~R.}\ \bibnamefont {Johnson}}, \bibinfo
  {author} {\bibfnamefont {C.~A.}\ \bibnamefont {Ryan}}, \ and\ \bibinfo
  {author} {\bibfnamefont {M.}~\bibnamefont {Steffen}},\ }\href@noop {}
  {\bibfield  {journal} {\bibinfo  {journal} {Phys. Rev. A}\ }\textbf {\bibinfo
  {volume} {87}},\ \bibinfo {pages} {062119} (\bibinfo {year}
  {2013})}\BibitemShut {NoStop}%
\bibitem [{\citenamefont {Blume-Kohout}\ \emph {et~al.}(2013)\citenamefont
  {Blume-Kohout}, \citenamefont {Gamble}, \citenamefont {Nielsen},
  \citenamefont {Mizrahi}, \citenamefont {Sterk},\ and\ \citenamefont
  {Maunz}}]{gst}%
  \BibitemOpen
  \bibfield  {author} {\bibinfo {author} {\bibfnamefont {R.}~\bibnamefont
  {Blume-Kohout}}, \bibinfo {author} {\bibfnamefont {J.~K.}\ \bibnamefont
  {Gamble}}, \bibinfo {author} {\bibfnamefont {E.}~\bibnamefont {Nielsen}},
  \bibinfo {author} {\bibfnamefont {J.}~\bibnamefont {Mizrahi}}, \bibinfo
  {author} {\bibfnamefont {J.~D.}\ \bibnamefont {Sterk}}, \ and\ \bibinfo
  {author} {\bibfnamefont {P.}~\bibnamefont {Maunz}},\ }\href@noop {}
  {\bibfield  {journal} {\bibinfo  {journal} {arXiv preprint arXiv:1310.4492}\
  } (\bibinfo {year} {2013})}\BibitemShut {NoStop}%
\bibitem [{\citenamefont {Stark}(2014)}]{stark}%
  \BibitemOpen
  \bibfield  {author} {\bibinfo {author} {\bibfnamefont {C.}~\bibnamefont
  {Stark}},\ }\href@noop {} {\bibfield  {journal} {\bibinfo  {journal} {Phys.
  Rev. A}\ }\textbf {\bibinfo {volume} {89}},\ \bibinfo {pages} {052109}
  (\bibinfo {year} {2014})}\BibitemShut {NoStop}%
\bibitem [{\citenamefont {Perez-Garcia}\ \emph {et~al.}(2006)\citenamefont
  {Perez-Garcia}, \citenamefont {Verstraete}, \citenamefont {Wolf},\ and\
  \citenamefont {Cirac}}]{perez2006matrix}%
  \BibitemOpen
  \bibfield  {author} {\bibinfo {author} {\bibfnamefont {D.}~\bibnamefont
  {Perez-Garcia}}, \bibinfo {author} {\bibfnamefont {F.}~\bibnamefont
  {Verstraete}}, \bibinfo {author} {\bibfnamefont {M.~M.}\ \bibnamefont
  {Wolf}}, \ and\ \bibinfo {author} {\bibfnamefont {J.~I.}\ \bibnamefont
  {Cirac}},\ }\href@noop {} {\bibfield  {journal} {\bibinfo  {journal} {arXiv
  preprint quant-ph/0608197}\ } (\bibinfo {year} {2006})}\BibitemShut {NoStop}%
\bibitem [{\citenamefont {Sch{\"o}n}\ \emph {et~al.}(2007)\citenamefont
  {Sch{\"o}n}, \citenamefont {Hammerer}, \citenamefont {Wolf}, \citenamefont
  {Cirac},\ and\ \citenamefont {Solano}}]{schon2007sequential}%
  \BibitemOpen
  \bibfield  {author} {\bibinfo {author} {\bibfnamefont {C.}~\bibnamefont
  {Sch{\"o}n}}, \bibinfo {author} {\bibfnamefont {K.}~\bibnamefont {Hammerer}},
  \bibinfo {author} {\bibfnamefont {M.~M.}\ \bibnamefont {Wolf}}, \bibinfo
  {author} {\bibfnamefont {J.~I.}\ \bibnamefont {Cirac}}, \ and\ \bibinfo
  {author} {\bibfnamefont {E.}~\bibnamefont {Solano}},\ }\href@noop {}
  {\bibfield  {journal} {\bibinfo  {journal} {Physical Review A}\ }\textbf
  {\bibinfo {volume} {75}},\ \bibinfo {pages} {032311} (\bibinfo {year}
  {2007})}\BibitemShut {NoStop}%
\bibitem [{\citenamefont {Crosswhite}\ and\ \citenamefont
  {Bacon}(2008)}]{crosswhite2008finite}%
  \BibitemOpen
  \bibfield  {author} {\bibinfo {author} {\bibfnamefont {G.~M.}\ \bibnamefont
  {Crosswhite}}\ and\ \bibinfo {author} {\bibfnamefont {D.}~\bibnamefont
  {Bacon}},\ }\href@noop {} {\bibfield  {journal} {\bibinfo  {journal}
  {Physical Review A}\ }\textbf {\bibinfo {volume} {78}},\ \bibinfo {pages}
  {012356} (\bibinfo {year} {2008})}\BibitemShut {NoStop}%
\bibitem [{\citenamefont {Geiger}\ \emph {et~al.}(2001)\citenamefont {Geiger},
  \citenamefont {Heckerman}, \citenamefont {King},\ and\ \citenamefont
  {Meek}}]{geiger2001stratified}%
  \BibitemOpen
  \bibfield  {author} {\bibinfo {author} {\bibfnamefont {D.}~\bibnamefont
  {Geiger}}, \bibinfo {author} {\bibfnamefont {D.}~\bibnamefont {Heckerman}},
  \bibinfo {author} {\bibfnamefont {H.}~\bibnamefont {King}}, \ and\ \bibinfo
  {author} {\bibfnamefont {C.}~\bibnamefont {Meek}},\ }\href@noop {} {\bibfield
   {journal} {\bibinfo  {journal} {Annals of statistics}\ ,\ \bibinfo {pages}
  {505}} (\bibinfo {year} {2001})}\BibitemShut {NoStop}%
\bibitem [{\citenamefont {Garcia}\ \emph {et~al.}(2005)\citenamefont {Garcia},
  \citenamefont {Stillman},\ and\ \citenamefont
  {Sturmfels}}]{garcia2005algebraic}%
  \BibitemOpen
  \bibfield  {author} {\bibinfo {author} {\bibfnamefont {L.~D.}\ \bibnamefont
  {Garcia}}, \bibinfo {author} {\bibfnamefont {M.}~\bibnamefont {Stillman}}, \
  and\ \bibinfo {author} {\bibfnamefont {B.}~\bibnamefont {Sturmfels}},\
  }\href@noop {} {\bibfield  {journal} {\bibinfo  {journal} {Journal of
  Symbolic Computation}\ }\textbf {\bibinfo {volume} {39}},\ \bibinfo {pages}
  {331} (\bibinfo {year} {2005})}\BibitemShut {NoStop}%
\bibitem [{\citenamefont {Henson}\ \emph {et~al.}(2014)\citenamefont {Henson},
  \citenamefont {Lal},\ and\ \citenamefont {Pusey}}]{henson2014theory}%
  \BibitemOpen
  \bibfield  {author} {\bibinfo {author} {\bibfnamefont {J.}~\bibnamefont
  {Henson}}, \bibinfo {author} {\bibfnamefont {R.}~\bibnamefont {Lal}}, \ and\
  \bibinfo {author} {\bibfnamefont {M.~F.}\ \bibnamefont {Pusey}},\ }\href@noop
  {} {\bibfield  {journal} {\bibinfo  {journal} {New Journal of Physics}\
  }\textbf {\bibinfo {volume} {16}},\ \bibinfo {pages} {113043} (\bibinfo
  {year} {2014})}\BibitemShut {NoStop}%
\bibitem [{\citenamefont {Hardy}(2001)}]{hardy2001quantum}%
  \BibitemOpen
  \bibfield  {author} {\bibinfo {author} {\bibfnamefont {L.}~\bibnamefont
  {Hardy}},\ }\href@noop {} {\bibfield  {journal} {\bibinfo  {journal} {arXiv
  preprint quant-ph/0101012}\ } (\bibinfo {year} {2001})}\BibitemShut {NoStop}%
\bibitem [{\citenamefont {Hardy}(2013)}]{hardy2013formalism}%
  \BibitemOpen
  \bibfield  {author} {\bibinfo {author} {\bibfnamefont {L.}~\bibnamefont
  {Hardy}},\ }\href@noop {} {\bibfield  {journal} {\bibinfo  {journal}
  {Mathematical Structures in Computer Science}\ }\textbf {\bibinfo {volume}
  {23}},\ \bibinfo {pages} {399} (\bibinfo {year} {2013})}\BibitemShut
  {NoStop}%
\bibitem [{Note1()}]{Note1}%
  \BibitemOpen
  \bibinfo {note} {The author is even inclined to suggest that states and
  outcomes should fundamentally be thought of as on an equal
  footing.}\BibitemShut {Stop}%
\bibitem [{Note2()}]{Note2}%
  \BibitemOpen
  \bibinfo {note} {One could parse correlations into two separate kinds of
  independence: The first kind being when ${\delimiter "426830A \protect
  \mathrm {Tr}\rho \Sigma \delimiter "526930B _a}^i = \protect \mathrm
  {Tr}{\delimiter "426830A \rho \delimiter "526930B _a}^i {\delimiter "426830A
  \Sigma \delimiter "526930B _a}^i$. The second kind being when ${\delimiter
  "426830A \rho \delimiter "526930B _a}^i = {\delimiter "426830A \rho
  \delimiter "526930B _a}$ and ${\delimiter "426830A \Sigma \delimiter "526930B
  _a}^i = {\delimiter "426830A \Sigma \delimiter "526930B }^i$}\BibitemShut
  {NoStop}%
\bibitem [{Note3()}]{Note3}%
  \BibitemOpen
  \bibinfo {note} {This second meaning of the word ``local'' should not be too
  confusing as it will be clear from context whether we are considering
  individual qudit observables or small numbers of state and measurement
  settings.}\BibitemShut {Stop}%
\bibitem [{Note4()}]{Note4}%
  \BibitemOpen
  \bibinfo {note} {Technically, we should write $(\sigma _q)_\mu $ to emphasize
  that the qudit measurements do not necessarily share a reference frame, but
  we will not write this here for the sake of reducing index
  clutter.}\BibitemShut {Stop}%
\bibitem [{Note5()}]{Note5}%
  \BibitemOpen
  \bibinfo {note} {If one considers data from $(d^2+1) \times (d^2+1)$ settings
  (as described in Appendix \ref {plus1},) then there are $\protect \binom
  {d^2+1}{2}^2$ distinct PDs, having already divided out the aforementioned
  equivalences. This is because one must choose the $d^2-1$ rows and $d^2-1$
  columns of the data that will be common to each corner. The remaining 2 rows
  and 2 columns are what actually displace the corners.}\BibitemShut {Stop}%
\bibitem [{\citenamefont {McCormick}\ \emph {et~al.}(2017)\citenamefont
  {McCormick}, \citenamefont {van Enk},\ and\ \citenamefont
  {Beck}}]{mccormick2017experimental}%
  \BibitemOpen
  \bibfield  {author} {\bibinfo {author} {\bibfnamefont {A.}~\bibnamefont
  {McCormick}}, \bibinfo {author} {\bibfnamefont {S.}~\bibnamefont {van Enk}},
  \ and\ \bibinfo {author} {\bibfnamefont {M.}~\bibnamefont {Beck}},\
  }\href@noop {} {\bibfield  {journal} {\bibinfo  {journal} {arXiv preprint
  arXiv:1701.06498}\ } (\bibinfo {year} {2017})}\BibitemShut {NoStop}%
\end{thebibliography}%
\end{document}